%% file: arxiv_mimicry.tex
\PassOptionsToPackage{table}{xcolor}
\documentclass[acmtog,nonacm]{acmart}
\pdfoutput=1

\AtBeginDocument{%
  \providecommand\BibTeX{{%
    \normalfont B\kern-0.5em{\scshape i\kern-0.25em b}\kern-0.8em\TeX}}}

\setcopyright{none}
\copyrightyear{2021}
\acmYear{2021}
\acmDOI{10.1145/1122445.1122456}
\renewcommand\footnotetextcopyrightpermission[1]{}
\acmBooktitle{}
\acmConference{}

\usepackage{booktabs}
\usepackage{enumitem}
\setlist{noitemsep}
\setlist[1]{labelindent=\parindent}
\setlist[itemize,1]{label=---}

\usepackage{ccicons}          
\usepackage{subcaption}
\usepackage{wrapfig}

\usepackage{epstopdf}
\usepackage{tikz-cd}

\DeclareGraphicsExtensions{%
    .pdf,.PDF,%
    .png,.PNG,%
    .jpg,.mps,.jpeg,.jbig2,.jb2,.JPG,.JPEG,.JBIG2,.JB2}

\graphicspath{{figures/}}

\newcommand{\BR}{\mathbb{R}}

\newcommand{\CM}{\mathcal{M}}
\newcommand{\CT}{\mathcal{T}}
\newcommand{\CX}{\mathcal{X}}
\newcommand{\CP}{\mathcal{P}}
\newcommand{\CQ}{\mathcal{Q}}

\newcommand{\Cc}{\mathsf{c}}
\newcommand{\Ch}{\mathsf{h}}

\newcommand{\bp}{\mathbf{p}}
\newcommand{\bq}{\mathbf{q}}
\newcommand{\br}{\mathbf{r}}

\newcommand{\bh}{\mathbf{h}}
\newcommand{\bc}{\mathbf{c}}
\newcommand{\bo}{\mathbf{o}}
\newcommand{\bd}{\mathbf{d}}
\newcommand{\be}{\mathbf{e}}
\newcommand{\bx}{\mathbf{x}}
\newcommand{\by}{\mathbf{y}}
\newcommand{\bz}{\mathbf{z}}
\newcommand{\bv}{\mathbf{v}}

\newcommand{\bC}{\mathbf{C}}

\newcommand{\HCQ}{\widehat{\mathcal{Q}}}
\newcommand{\Hbq}{\widehat{\mathbf{q}}}

\DeclareMathOperator*{\argmin}{arg\,min}

\DeclareMathOperator*{\bary}{Bary}

\definecolor{myorchid}{RGB}{10,10,200}
\definecolor{darkgreen}{RGB}{10,150,10}

\definecolor{tableblue}{RGB}{49,130,189}
\definecolor{tablered}{RGB}{222,45,38}
\definecolor{tablegreen}{RGB}{49,163,84}

\definecolor{matlabblue}{rgb}{0,0.447,0.741}
\definecolor{matlabred}{rgb}{0.85,0.325,0.098}
\definecolor{matlaborange}{rgb}{1.0,0.51,0.0}
\definecolor{matlabpurple}{rgb}{0.494,0.184,0.556}

\definecolor{participant}{RGB}{90,90,90}

\newcommand{\comm}[3]{}

\newcommand{\revm}[1] {#1}
\newcommand{\rev}[1] {#1}

\makeatletter
\newcommand{\pquote}[2][\@nil]{
    \def\tmp{#1}%
    \ifx\tmp\@nnil
        \textcolor{participant}{{``#2''}}
    \else
        \textcolor{participant}{{``#2'' \emph{(#1)}}}
    \fi
}

\setcitestyle{square}

\begin{document}

\title{Mid-Air Drawing of Curves on 3D Surfaces in Virtual Reality}

\author{Rahul Arora}
\email{arorar@dgp.toronto.edu}
\orcid{0000-0001-7281-8117}
\affiliation{%
  \institution{University of Toronto}
  \city{Toronto}
  \state{Ontario}
  \country{Canada}
}

\author{Karan Singh}
\email{karan@dgp.toronto.edu}
\affiliation{%
  \institution{University of Toronto}
  \city{Toronto}
  \state{Ontario}
  \country{Canada}
}

\renewcommand{\shortauthors}{Arora and Singh}

\begin{abstract}

Complex 3D curves can be created by directly drawing mid-air in immersive environments (\revm{Augmented and Virtual Realities}). Drawing mid-air strokes precisely on the surface of a 3D virtual object\rev{,} however, is difficult; necessitating a projection of the mid-air stroke onto the user ``intended'' surface curve. We present the first detailed investigation of the fundamental problem of 3D stroke projection in \revm{VR}. An assessment of the design requirements of real-time drawing of curves on 3D objects in \revm{VR} is followed by the definition and classification of multiple techniques for 3D stroke projection. We analyze the advantages and shortcomings of these approaches both theoretically and via practical pilot testing. We then formally evaluate the two most promising techniques \emph{spraycan} and \emph{mimicry} with 20 users in VR. The study shows a strong qualitative and quantitative user preference for our novel stroke \emph{mimicry} projection algorithm. We further illustrate the effectiveness and utility of stroke mimicry, to draw complex 3D curves on surfaces for various artistic and functional design applications.

\end{abstract}



\begin{teaserfigure}
    \centering
    \includegraphics[width=\linewidth]{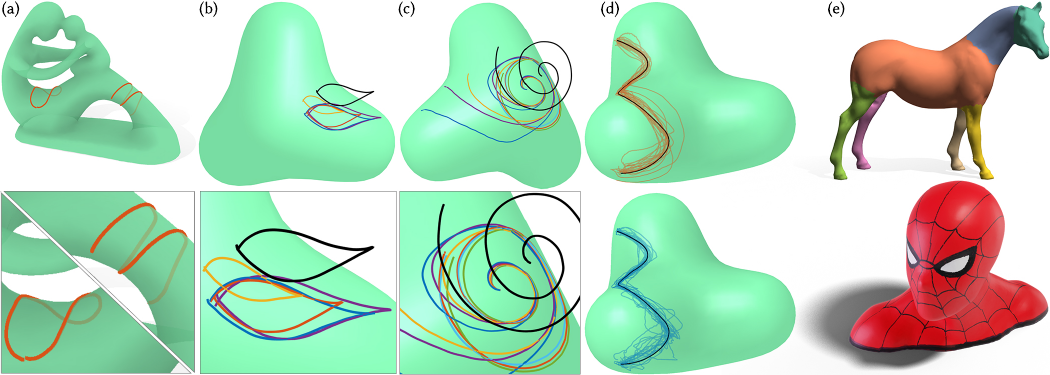}
    \caption{Drawing curves mid-air that lie precisely on the surface of a virtual 3D object in AR/VR is difficult (a). Projecting mid-air 3D strokes (\textbf{black}) onto 3D objects is an under-constrained problem with many seemingly reasonable solutions (b). We analyze this fundamental AR/VR problem of 3D stroke projection, define and characterize multiple novel projection techniques (c), and test the two most promising approaches---\emph{spraycan} shown in \textbf{\textcolor{matlabblue}{blue}} and \emph{mimicry} shown in \textbf{\textcolor{matlabred}{red}} in (b)--(d)---using a quantitative study with 20 users (d). The user-preferred \emph{mimicry} technique attempts to mimic the 3D mid-air stroke as closely as possible when projecting onto the virtual object. We showcase the importance of drawing curves on 3D surfaces, and the utility of our novel \emph{mimicry} approach, using multiple artistic and functional applications (e) such as interactive shape segmentation (top) and texture painting (bottom). }
    \label{fig:teaser}
\end{teaserfigure}

\maketitle

\section{Introduction}
\label{sec:intro}
\input{1_intro}

\section{Related Work}
\label{sec:related}

\input{2_related}

\section{Projecting Strokes on 3D Objects}
\label{sec:drawing}

\input{3_drawing}

\section{Anchored Stroke Projection}
\label{sec:historical}
\input{4_historical}

\section{User Study}
\label{sec:study}
\input{5_study}

\section{Study Results and Discussion}
\label{sec:results}
\input{6_results}


\section{Applications}
\label{sec:applications}
\input{7_applications}

\section{Conclusion}
\label{sec:conclusion}
\input{8_conclude}


\bibliographystyle{ACM-Reference-Format}
\bibliography{references}

\input{99_appendix}

\end{document}

%% file: 1_intro.tex

\begin{figure}
    \centering
    \includegraphics[width=\linewidth]{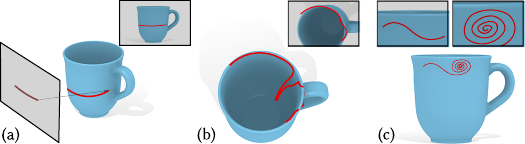}
    \caption{Stroke projection using a 2D interface is typically WYSIWYG: 2D points along a user stroke (a, inset) are ray-cast through the given view to create corresponding 3D curve points on the surface of 3D scene objects (a). Even small errors or noise in 2D strokes can cause large discontinuities in 3D, especially near ridges and sharp features (b). Complex curves spanning many viewpoints, or with large scale variations in detail, often require the curve to be drawn in segments from multiple user-adjusted viewpoints (c).}
    \label{fig:2DProblems}
\end{figure}

Drawing is a fundamental tool of human visual expression and communication. 
Digital sketching with pens, styli, mice, and even fingers in 2D is ubiquitous in visually creative computing applications. Drawing or painting \textbf{on} 3D virtual objects for example, is critical to interactive 3D modelling, animation, and visualization, where its uses include: object selection, annotation, and segmentation \cite{heckel2013sketch,meng2011icutter,jung2002annotating}; 3D curve and surface design \cite{igarashi1999teddy,nealen2007fibermesh}; strokes for 3D model texturing or painterly rendering \cite{kalnins2002wysiwyg} (Figure~\ref{fig:teaser}e). 
In 2D, digitally drawn on-screen strokes are WYSIWYG mapped onto 3D virtual objects, by projecting 2D stroke points through the given view onto the virtual object(s) (Figure~\ref{fig:2DProblems}a).

\begin{figure}
    \centering
    \includegraphics[width=\linewidth]{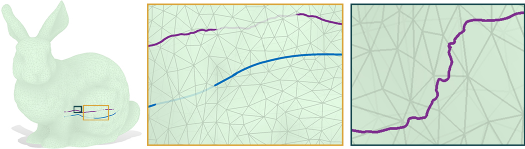}
    \caption{Mid-air drawing precisely on a 3D virtual object is difficult (faint regions of strokes are \rev{behind} the surface), regardless of drawing quick smooth strokes (\textbf{\textcolor{matlabblue}{blue}}), or slow detailed strokes (\textbf{\textcolor{matlabpurple}{purple}}). Deliberately slow drawing is further detrimental to stroke aesthetic (right).}
    \label{fig:imprecise3D}
\end{figure}

Sketching in immersive environments (AR/VR) has the mystical aura of a magical wand, allowing users to draw directly in 3D. Mid-air drawing has the potential to significantly disrupt interactive 3D graphics, as evidenced by the increasing popularity of applications such as Tilt Brush~\cite{google2020tilt} and Quill~\cite{oculus2020quill}.
A fundamental requirement for numerous interactive 3D applications in AR/VR \rev{is} the ability to directly draw, or project drawn 3D strokes, precisely on virtual objects.
While directly drawing on a physical object is reasonably easy, drawing directly on a virtual 3D object is near impossible without haptic constraints (Figure~\ref{fig:imprecise3D}). Furthermore, unlike 2D drawing, where the WYSIWYG view-based projection of 2D strokes onto 3D objects is unambiguously clear, the user-intended mapping of a mid-air 3D stroke onto a 3D object is less obvious. We present the first detailed investigation into plausible user-intended projections of mid-air strokes on to 3D virtual objects.

Interfaces for 2D/3D curve creation in general, use perceptual insights or geometric assumptions like smoothness and planarity, to project, neaten, or otherwise process sketched strokes. Some applications wait for user stroke completion before processing it in entirety, for example when fitting splines \cite{bae08ilovesketch}.
Our goal is to establish an application agnostic, base-line projection approach for mid-air 3D strokes. We thus assume a stroke is processed while being drawn and inked in real-time, i.e., the output curve corresponding to a partially drawn stroke is fixed/inked in real-time, based on partial stroke input \cite{thiel2011elasticurves}. 

One might further conjecture that all ``reasonable'' and mostly continuous projections would produce similar results, as long as users are given interactive visual feedback of the projection. This is indeed true for tasks requiring discrete point-on-surface selection, where users can freely re-position the drawing tool until its interactively visible projection corresponds to user-intent.
Real-time curve drawing, however, is very sensitive to the projection technique, where any mismatch between user intention and algorithmic projection, is continuously inked into the projected curve (Figure~\ref{fig:teaser}d).

\paragraph{2D Strokes Projected onto 3D Objects} The standard user-intended mapping of a 2D on-screen stroke is a raycast projection through the given monocular viewpoint. Raycasting is WYSIWYG (What You See Is What You Get): the 3D curve visually matches the 2D stroke from said viewpoint (Figure~\ref{fig:2DProblems}a). Ongoing research on mapping 2D strokes to 3D objects assumes this fundamental \emph{view-centric} projection, focusing instead on specific problems such as creating curves around ridge/valley features (where small 2D error can cause large 3D depth error, Figure~\ref{fig:2DProblems}b); or drawing complex curves with large scale variation (where multiple viewpoint changes are needed while drawing, Figure~\ref{fig:2DProblems}c).
These problems are mitigated by the direct 3D input and viewing flexibility of AR/VR, assuming the mid-air stroke to 3D object projection matches user intent.

\paragraph{3D Strokes Projected onto 3D Objects} Physical analogies motivate existing approaches to defining a user-intended projection from 3D points in a mid-air stroke to 3D points on a virtual object (Figure~\ref{fig:historyFree}). Graffiti-style painting with a \emph{spraycan} is arguably the current standard, deployed in commercial immersive paint and sculpt software such as \rev{Medium~\cite{adobe2021medium}} and Gravity Sketch~\shortcite{gravity2020gravity}.
A closest-point projection approximates drawing with the tool on the 3D object, without actual physical contact (used by the "guides" tool in Tilt Brush \rev{\cite{google2020tilt}}). Like view-centric 2D stroke projection, these approaches are \emph{context-free}: processing each mid-air point independently.
The AR/VR drawing environment comprising six--degree of freedom controller input and unconstrained binocular viewing, is however, significantly richer than 2D sketching. The user-intended projection of a mid-air stroke (\S~\ref{sec:drawing}) as a result is complex, influenced by the ever-changing 3D relationship between the view, drawing controller and virtual object. We therefore argue the need for historical context (i.e., the partially drawn stroke and its projection) in determining the projection of a given stroke point.
We balance the use of this historical context, with the overarching goal of a general purpose projection that makes little or no assumption on the nature of the user stroke or its projection.

We thus explore \emph{anchored} projection techniques, that minimally use the most recently projected stroke point, as context for projecting the current stroke point (\S~\ref{sec:historical}). We evaluate various anchored projections, both theoretically and practically by pilot testing. Our most promising and novel approach \emph{anchored-smooth-closest-point} (also called \emph{mimicry}), captures the natural tendency of a user stroke to mimic the shape of the desired projected curve. A formal user study \rev{in VR} (\S~\ref{sec:study}) shows \emph{mimicry} to perform significantly better than \emph{spraycan} (the current baseline) in producing curves that match user intent (\S~\ref{sec:results}). \rev{While our formal evaluation is limited to VR, the fundamental problem we study \revm{could directly translate} to AR scenarios as well.} This paper thus contributes, to the best of our knowledge, the first principled investigation of real-time inked techniques to project 3D mid-air strokes drawn in \revm{VR} onto 3D virtual objects, and a novel stroke projection benchmark for \revm{VR}: \emph{mimicry}.


%% file: 2_related.tex
Our work is related to research on drawing and sculpting in immersive realities, interfaces for drawing curves on, near, and around surfaces, and sketch-based modelling tools.

\subsection{Immersive Sketching and Modelling}
\label{sec:relatedDrawing}

Immersive creation has a long history in computer graphics. Immersive 3D sketching was pioneered by the HoloSketch system \cite{deering1995holosketch}, which used a 6-DoF wand as the input device for creating polyline sketches, 3D tubes, and primitives. In a similar vein, various subsequent systems have explored the creation of freeform 3D curves and swept surfaces \cite{schkolne2001surface, keefe2001cavepainting, google2020tilt}.
While directly turning 3D input to creative output is acceptable for ideation, the inherent imprecision of 3D sketching is quickly apparent when more structured creation is desired.

The perceptual and ergonomic challenges in precise control of 3D input is well-known~\cite{keefe2007drawing, wiese2010investigating, arora2017experimental, machuca2019effect, machuca2018multi}, resulting in various methods for correcting 3D input.
Input 3D curves have been algorithmically regularized to snap onto existing geometry, as with the FreeDrawer \rev{\cite{wesche2001freedrawer}} system, or constrained physically to 2D input with additional techniques for ``lifting'' these curves into 3D \cite{jackson2016lift, arora2018symbiosis, kwan2019mobi, paczkowski2011insitu}. Haptic rendering devices \cite{keefe2007drawing, kamuro20113d} and tools utilizing passive physical feedback \cite{grossman2002creating} are an alternate approach to tackling the imprecision of 3D inputs. We are motivated by similar considerations.

Arora et al.~\shortcite{arora2017experimental} demonstrated the difficulty of creating curves that lie exactly on virtual surfaces in VR, even when the virtual surface is a plane. This observation directly motivates our exploration of techniques for projecting 3D strokes onto surfaces, instead \revm{of coercing} users to awkwardly draw exactly on a virtual surface. 

\subsection{Drawing Curves on, near, and around Surfaces}
\label{sec:relatedProjecting}

Curve creation and editing on or near the surface of 3D virtual objects is fundamental for a variety of artistic and functional shape modelling tasks. 
Functionally, curves on 3D surfaces are used to model or annotate structural features \cite{iwires, neobarok}, define trims and holes \cite{schmidt2010meshmixer}, and to provide handles for shape deformation \cite{wires, kara2007sketch,nealen2007fibermesh}, registration \cite{gehre2018interactive} and remeshing \cite{krishlevoy,takayama2013sketch}.
Artistically, curves on surfaces are used in painterly rendering \cite{goochbook}, decal creation \cite{schmidt2006discrete}, texture painting \cite{adobe2020substance}, and even texture synthesis \cite{fisher2007design}. Curve on surface creation in this body of research typically uses the established view-centric WYSIWYG projection of on-screen sketched 2D strokes. While the sketch view-point in these interfaces is interactively set by the user, there has been some effort in automatic camera control for drawing \cite{ortega2014direct}, auto-rotation of the sketching view for 3D planar curves \cite{flatfab}, and user assistance in selecting the most \emph{sketchable} viewpoints \cite{bae08ilovesketch}. Immersive 3D drawing enables direct, view-point independent 3D curve sketching, and is thus an appealing alternative to these 2D interfaces.

Our work is also related to drawing curves \emph{around} surfaces. Such techniques are important for a variety of applications: modelling string and wire that wrap around objects \cite{coleman2006cords}; curves that loosely conform to virtual objects \cite{krs2017Skippy}; clothing design on a 3D mannequin \cite{turquin2007sketch}; layered modelling of shells and armour \cite{depauli2015secondskin}; and the design and grooming of hair and fur \cite{fu2007sketching,schmidt2011overcoat, xing2019hairbrush}. Some approaches such as SecondSkin \rev{\cite{ depauli2015secondskin}} and Skippy \rev{\cite{krs2017Skippy}} use insights into spatial relationship between a 2D stroke and the 3D object, to infer a 3D curve that lies on and around the surface of the object.  Other techniques like Cords \rev{\cite{coleman2006cords}} or hair and clothing design \cite{xing2019hairbrush} are closer to our work, in that they drape 3D curve input on and around 3D objects using geometric collisions or physical simulation.
In contrast, this paper is focused on the general problem of projecting a drawn 3D stroke to a real-time inked curve on the surface of a 3D object. While we do not address curve creation with specific geometric relationships to the object surface (like distance-offset curve), our techniques can be extended to incorporate geometry-specific terms (\S~\ref{sec:conclusion}).

\subsection{Sketch-based 3D Modelling}
\label{sec:relatedModeling}

Sketch-based 3D modelling is a rich ongoing area of research (see survey by Olsen et al.~\shortcite{olsen2009sketch}). Typically, these systems interpret 2D sketch inputs for various shape modelling tasks. One could categorize these modelling approaches as single-view (akin to traditional pen on paper) \cite{schmidt2009analytic,andre11single,chen20133sweep,xu2014true2form} or multi-view (akin to 3D modelling with frequent view manipulation) \cite{igarashi1999teddy,fan2004sketch,nealen2007fibermesh,bae08ilovesketch,fan2013modeling}. Single-view techniques use perceptual insights and geometric properties of the 2D sketch to infer its depth in 3D, while multi-view techniques explicitly use view manipulation to specify 3D curve attributes from different views. While our work utilizes mid-air 3D stroke input, the ambiguity of projection onto surfaces connects it to the interpretative algorithms designed for sketch-based 3D modelling. 
We aim to take advantage of the immersive interaction space by allowing view manipulation as and when desired, independent of geometry creation.

%% file: 3_drawing.tex
We first formally state the problem of projecting a mid-air 3D stroke onto a 3D virtual object.
Let $\CM=(V, E, F)$ be a 3D object, represented as a manifold triangle mesh embedded in $\BR^3$.
A user draws a piece-wise linear mid-air stroke by moving a 6-DoF controller or drawing tool in \revm{VR}. The 3D stroke $\CP \subset \BR^3$ is a sequence of $n$ points $(\bp_i)_{i=0}^{n-1}$, connected by line segments. Corresponding to each point $\bp_i\in \BR^3$, is a system state $S_i = (\bh_i, \bc_i, \Ch_i, \Cc_i)$, where $\bh_i, \bc_i \in \BR^3$ are the positions of the headset and the controller, respectively, and $\Ch_i, \Cc_i \in Sp(1)$ are their respective orientations, represented as unit quaternions.
Also, without loss of generality, assume $\bc_i = \bp_i$, i.e. the controller positions describe the stroke points $\bp_i$.

We want to define a \emph{projection} $\pi$, which transforms the sequence of points $(\bp_i)_{i=0}^{n-1}$ to a corresponding sequence of points  $(\bq_i)_{i=0}^{n-1}$ on the 3D virtual object, i.e. $\bq_i\in \CM$.
Consecutive points in this sequence are connected by geodesics on $\CM$, \rev{describing} the \emph{projected curve} $\CQ\subset \CM$.
The aim of a successful projection method of course, is to match the undisclosed user-intended curve. The projection is also designed for real-time inking of curves: points $\bp_i$ are processed upon input and projected in real-time (under 100ms) to $\bq_i$ using the current system state \rev{$S_i$}, and optionally, prior system states $(S_j)_{j=0}^{i-1}$, stroke points $(\bp_j)_{j=0}^{i-1}$ and projections $(\bq_j)_{j=0}^{i-1}$.


\begin{figure}
    \centering
    \includegraphics[width=\linewidth]{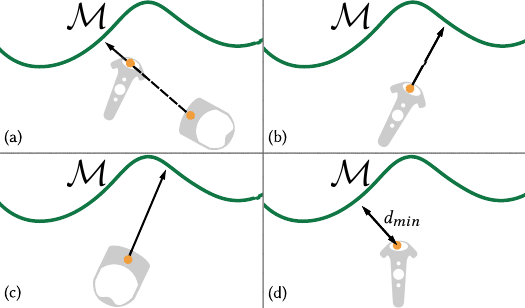}
    \caption{Context-free techniques: \emph{occlude} projects points from the controller origin along the direction from the  eye (HMD origin) to the controller (a); \emph{spraycan} projects \revm{points} from the controller origin in a direction defined by the controller's orientation (b); \emph{head-centric}, akin to 2D projects points along the view direction defined by HMD orientation (c); \emph{snap} projects points from the controller origin to their closest-point on $\CM$ (d).}
    \label{fig:historyFree}
\end{figure}

\subsection{Context-Free Projection Techniques}
\label{sec:contextFree}

Context-free techniques project points independent of each other, simply based on the spatial relationships between the controller, HMD, and 3D object ( Figure~\ref{fig:historyFree}). 
We can further categorize techniques as raycast or proximity based.

\subsubsection{Raycast Projections}
\label{sec:raycasting}

View-centric projection in 2D interfaces \rev{projects} points from the screen along a ray from the eye through the screen point, to where the ray first intersects the 3D object. In an immersive setting, raycast approaches similarly use a ray emanating from the 3D stroke point to intersect 3D objects. This ray $(\bo, \bd)$ with origin $\bo$ and direction $\bd$ can be defined in a number of ways. Similar to pointing behaviour, \emph{occlude} defines this ray from the eye through the controller origin (Figure~\ref{fig:historyFree}a) $\left(\bc_i, (\bc_i - \bh_i) / \|\bc_i - \bh_i\|\right)$. If the ray intersects $\CM$, then the closest intersection to $\bp_i$ defines $\bq_i$. In case of no intersection, $\bp_i$ is ignored in defining the projected curve, i.e., $\bq_i$ is marked undefined and the projected curve connects $\bq_{i-1}$ to $\bq_{i+1}$ (or the proximal index points on either side of $i$ for which projections are defined). 
The \emph{spraycan} approach treats the controller like a spraycan, defining the ray like a nozzle direction in the local space of the controller (Figure~\ref{fig:historyFree}b). For example the ray could be defined as $(\bc_i, \mathbf{f}_i)$, where the nozzle $\mathbf{f}_i = \Cc_i\cdot [0, 0, 1]^T$ is the controller's local z-axis (or \emph{forward} direction).
Alternately, \emph{head-centric} projection can define the ray using the HMD's view direction as $(\bh_i, \Ch_i\cdot [0, 0, 1]^T)$ (Figure~\ref{fig:historyFree}c).

\begin{figure}
    \centering
    \includegraphics[width=\linewidth]{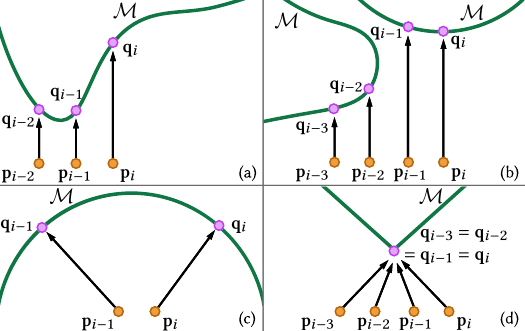}
    \caption{Context-free projection problems: \revm{large depth disparity (a), unexpected jumps (b), projection discontinuities (c), and undesirable snapping (d).}}
    \label{fig:contextFreeProblems}
\end{figure}

{\bf Pros and Cons:} The strengths of raycasting are: a predictable visual/proprioceptive sense of ray direction; a spatially continuous mapping between user input and projection rays; and scenarios where it is difficult or undesirable to reach and draw close to the virtual object.
Its biggest limitation stems from the controller/HMD-based ray direction being completely agnostic of the shape or location of the 3D object. Projected curves can consequently be very different in shape and size from drawn strokes \rev{(Figure~\ref{fig:contextFreeProblems}a--b)}, and ill-defined for stroke points with no ray-object intersection.

\subsubsection{Proximity-Based Projections}
\label{sec:proximity}

In 2D interfaces, the on-screen 2D strokes are typically distant to the viewed 3D scene, necessitating some form of raycast projection onto the visible surface of 3D objects. In AR/VR, however, users are able to reach out in 3D and directly draw the desired curve on the 3D object.
While precise mid-air drawing on a virtual surface is very difficult in practice (Figure~\ref{fig:imprecise3D}), projection methods based on proximity between the mid-air stroke and the 3D object are certainly worth investigation.

The simplest proximity-based projection technique \emph{snap}, projects a stroke point $\bp_i$ to its closest-point in $\CM$ (Figure~\ref{fig:historyFree}d).
\begin{equation}
    \bq_i = \pi_{snap}(\bp_i) = \argmin_{\bx\in\CM}\ d(\bp_i, \bx),
\end{equation}
\begin{wrapfigure}[10]{r}{0.3\columnwidth}
  \begin{center}
  		\vspace{-0.7cm}
  		\hspace{-0.7cm}
      \includegraphics[width=\linewidth]{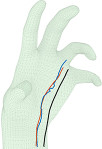}
  \end{center}
\end{wrapfigure}
where $d(\cdot, \cdot)$ is the Euclidean distance between two points.
Unfortunately, for triangle meshes, closest-point projection tends to snap to \rev{mesh edges} (\textcolor{matlabblue}{\textbf{blue}} curve inset), resulting in unexpectedly jaggy projected curves, even for smooth 3D input strokes (\textbf{black} curve inset)~\cite{panozzo2013weighted}.
These discontinuities are due to the discrete nature of the mesh representation, as well \rev{as} spatial singularities in closest point computation even for smooth 3D  objects. We mitigate this problem by formulating an extension of Panozzo et al.'s \emph{Phong} projection~\shortcite{panozzo2013weighted} in \S~\ref{sec:phong}, that simulates projection onto an imaginary smooth surface approximated by the mesh.
We denote this \emph{smooth-closest-point} projection as $\pi_{SCP}$ (\textcolor{matlabred}{\textbf{red}} curve inset).

{\bf Pros and Cons:} The biggest strength of proximity-based projection is it exploits the immersive concept of drawing directly on or near an object, using the spatial relationship between a 3D stroke point and the 3D object to determine projection. The main limitation is that since users rarely draw precisely on the surface, \rev{discontinuities in concave regions (Figure~\ref{fig:contextFreeProblems}c) and undesirable snapping in highly-convex regions 
(Figure~\ref{fig:contextFreeProblems}d)} persist when projecting distantly drawn stoke points, even when using \emph{smooth-closest-point}. In \S~\ref{sec:anchoredProximity}, we address this problem using stroke \emph{mimicry} to anchor distant stroke points close to the object to be finally projected using \emph{smooth-closest-point}.

\subsection{Smooth-Closest-Point Projection}
\label{sec:phong}

\tikzcdset{column sep/huge=3.5cm}
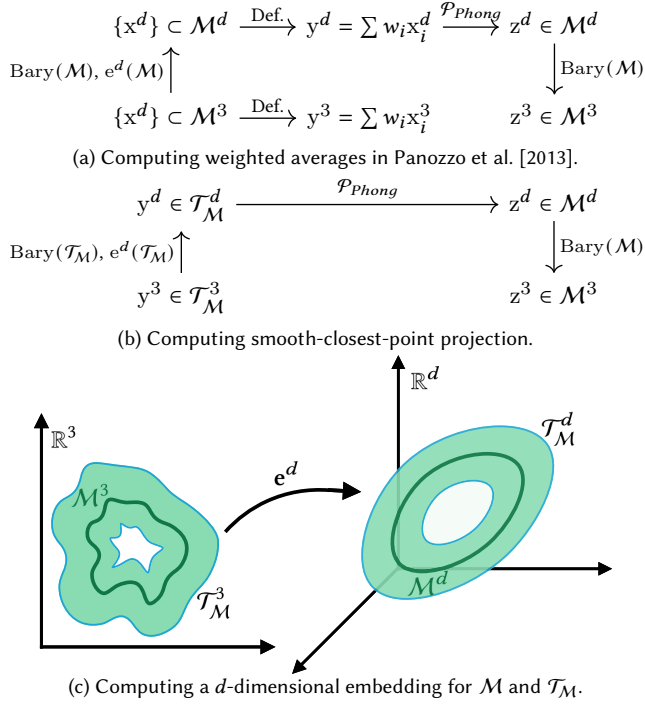
\begin{figure}
\centering
    \begin{subfigure}{\linewidth}
        \begin{tikzcd}
            \{\bx^d\}\subset\CM^d \arrow[r, "\text{Def.}"] & \by^d=\sum w_i\bx_i^d \arrow[r, "\CP_{Phong}"] & \bz^d\in\CM^d \arrow[d, "\bary(\CM)"] \\
            \{\bx^d\}\subset\CM^3 \arrow[u, "{\bary(\CM),\ \be^d(\CM)}"] \arrow[r, "\text{Def.}"] & \by^3=\sum w_i\bx_i^3                            & \bz^3\in\CM^3
        \end{tikzcd}
		\vspace{-.2cm}
        \caption{Computing weighted averages in Panozzo et al.~\shortcite{panozzo2013weighted}.}
    \end{subfigure}
    \begin{subfigure}{\linewidth}
        \begin{tikzcd}[column sep = huge]
            \by^d\in\CT_\CM^d \arrow[r, "\CP_{Phong}"] & \bz^d\in\CM^d \arrow[d, "\bary(\CM)"] \\
            \by^3\in\CT_\CM^3 \arrow[u, "{\bary(\CT_\CM),\ \be^d(\CT_\CM)}"] & \bz^3\in\CM^3
        \end{tikzcd}
		\vspace{-.2cm}
        \caption{Computing smooth-closest-point projection.}
    \end{subfigure}
    \begin{subfigure}{\linewidth}
    \centering
        \includegraphics[width=.9\linewidth]{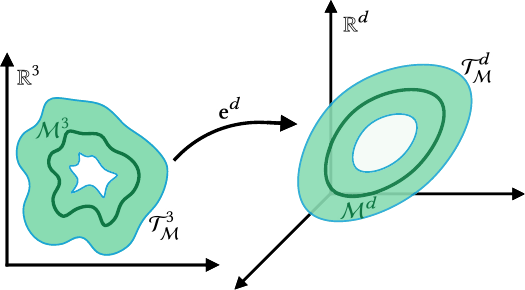}
		\vspace{-.2cm}
        \caption{Computing a $d$-dimensional embedding for $\CM$ and $\CT_\CM$.}
    \end{subfigure}
\caption{Panozzo et al.~\shortcite{panozzo2013weighted} compute weighted averages on surfaces (a), while we want to compute a smooth closest-point projection for an arbitrary point \emph{near} the mesh in $\BR^3$ (b). We therefore embed $\CT_\CM$---the region \emph{around} the mesh---in higher-dimensional space $\BR^d$, instead of just $\CM$ (c).}
\label{fig:phongAdapt}
\end{figure}

Our goal with \emph{smooth-closest-point} projection is to define a mapping from a 3D point to a point on $\CM$ that approximates the closest point projection but tends to be functionally smooth, at least for points near the 3D object. We note that computing the closest point to a Laplacian-smoothed mesh proxy, for example, will also provide a smoother mapping than $\pi_{snap}$, but a potentially poor closest-point approximation to the original mesh.

\emph{Phong projection}, introduced by Panozzo et al.~\shortcite{panozzo2013weighted}, addresses these goals for points expressible as weighted-averages of points on $\CM$, but we extend their technique to define a smooth-closest-point projection for points in the neighbourhood of the mesh. For completeness, we first present a brief overview of their technique.

Phong projection is a two-step approach to map a point $\by^3 \in \BR^3$ to a manifold triangle mesh $\CM$ embedded in $\BR^3$, emulating closest-point projection on a smooth surface approximated by the triangle mesh.
First, $\CM$ is embedded in a higher dimensional Euclidean space $\BR^d$ such that Euclidean distance (between points on the mesh) in $\BR^d$ approximates geodesic distances in $\BR^3$.
Second, analogous to vertex normal interpolation in Phong shading, a smooth surface is approximated by blending tangent planes across edges. Barycentric coordinates at a point within a triangle are used to blend the tangent planes corresponding to the three edges incident to the triangle.
We extend the first step to a higher dimensional embedding of not just the triangle mesh $\CM$, but a tetrahedral \rev{mesh} of an offset volume around the mesh $\CM$ (Figure~\ref{fig:phongAdapt}).
The second step remains the same, and we refer the reader to Panozzo et al.~\shortcite{panozzo2013weighted} for details. \rev{Such offset volumes, or \textit{shells}, around triangle meshes have also been utilized in recent methods for curve design~\cite{jin2019shell} and for attribute transfer between similar triangle meshes~\cite{jiang2020bijective}}.

For clarity, we refer to $\CM$ embedded in $\BR^3$ as $\CM^3$, and the embedding in $\BR^d$ as $\CM^d$.
Panozzo et al. compute $\CM^d$ by first embedding a subset of the vertices in $\BR^D$ using metric multi-dimensional scaling (MDS)~\cite{cox2008multidimensional}, aiming to preserve the geodesic distance between the vertices.
The embedding of the remaining vertices is then computed using LS-meshes~\cite{sorkine2004LSmeshes}.

For the problem of computing weighted averages on surfaces, one only needs to project 3D points of the form $\by^3=\sum w_i\bx_i^3$, where all $\bx_i^3\in\CM^3$.
The point $\by^3$ is \emph{lifted} into $\BR^d$ by simply defining $\by^d = \sum w_i\bx_i^d$, where $\bx_i^d$ is defined as the point on $\CM^d$ with the same implicit coordinates (triangle and barycentric coordinates) as $\bx_i^3$ does on $\CM^3$.
Therefore, their approach only embeds $\CM$ into $\BR^d$ (Figure~\ref{fig:phongAdapt}a,c).
In contrast, we want to project arbitrary points \emph{near} $\CM^3$ onto it using the Phong projection.
Therefore, we compute the offset surfaces at signed-distance $\pm\mu$ from $\CM$.
We then compute a tetrahedral mesh $\CT_\CM^3$ of the space between these two surfaces in $\BR^3$.
\rev{In the final precomputation step}, we embed the vertices of $\CT_\CM$ in $\BR^d$ using MDS and LS-Meshes as described above.

Now, given a 3D point $\by^3$ within a distance $\mu$ from $\CM^3$, we situate it within $\CT_\CM^3$, use tetrahedral Barycentric coordinates to infer its location in $\BR^d$, and then compute its Phong projection (Figure~\ref{fig:phongAdapt}b,c).
We fallback to closest-point projection for points outside $\CT_\CM^3$, since Phong projection converges to closest-point projection when far from $\CM$. Furthermore, we set $\mu$ large enough to easily handle our smooth-closest-point queries in \S~\ref{sec:anchoredProximity}.

\subsubsection{\rev{Projection Quality and Robustness Tests}}
\rev{Since the desirable properties of the Phong projection are not theoretically guaranteed for shapes with sharp features and noisy meshes~\cite{panozzo2013weighted}, we experimentally measure the quality of the embedding by testing for extreme dihedral angles---below 5\textdegree\ or above 175\textdegree---resulting in sliver tets in the $\BR^d$-embedding (Table~\ref{tbl:phongQuality}). Further, for a direct measure of projection quality, we densely sampled points in $\CT_\CM$ (four points per tet) and projected each using both $\pi_{snap}$ as well as $\pi_{SCP}$. Typically, we expect $\pi_{SCP}$ to be a smoother version of $\pi_{snap}$ (\S~\ref{sec:proximity}). Therefore, a $\pi_{SCP}$ projection much farther from the input than $\pi_{snap}$ indicates a clear failure:}

\vspace{-0.15cm}
\begin{equation}
\rev{\|\bp - \pi_{SCP}(\bp)\| > \|\bp - \pi_{snap}(\bp)\| + \|\mathtt{BBox}(\CM^3)\|/20\text{,}}
\label{eq:phongFailure}
\end{equation}

\rev{where $\|\mathtt{BBox}(\CM^3)\|$ is the length of the bounding box diagonal of $\CM^3$. Table~\ref{tbl:phongQuality} shows that the projection works well for almost all the sampled points. We also practically tested all the shapes by drawing myriad curves on each, but did not notice any clear failures of $\pi_{SCP}$. Finally, we stress-tested the technique using noisy versions of the unit cube mesh. At extreme levels of noise, when each vertex is moved in the normal direction by up to 20\% of the cube size, some clear failures showed up (Figure~\ref{fig:stressTests}d). In practice, such failures can be detected heuristically and $\pi_{snap}$ can be a drop-in replacement for such points. We, however, did not implement such a fix for our user study, or for the results shown in the paper.}

\setlength{\tabcolsep}{3pt}
\begin{table}
\centering
\caption{\rev{Embedding quality and $\pi_{SCP}$ failure results. The former is indicated by the percentage of dihedral angles <5\textdegree\ or >175\textdegree\ in the $\BR^d$-embedding, and the latter is defined in Eq.~\ref{eq:phongFailure} (lower values are desirable). Also shown are mesh sizes: (\#vertices, \#faces) for $\CM$ and (\#vertices, \#tets) for $\CT_\CM$.}}
\label{tbl:phongQuality}
\rowcolors{2}{white}{gray!25}
\resizebox{1.\linewidth}{!}{
\begin{tabular}{lrrrr}
\toprule
\textbf{Shape} & $|\CM|$ & $|\CT_\CM|$ & \textbf{\% slivers} & \textbf{\% $\pi_{SCP}$ fail} \\
\midrule
Trebol & (1.2K, 2.3K) & (7.9K, 38K) & $6.89$ & $0.00$\\
Cube & (1.5K, 3.0K) & (5.7K, 26K) & $17.44$ & $<0.01$\\
Torus & (1.7K, 3.5K) & (11K, 58K) & $0.41$ & $0.00$\\
Spiderman & (3.3K, 6.6K) & (15K, 83K) & $2.01$ & $0.02$\\
Hand & (4.2K, 8.5K) & (19K, 114K) & $0.49$ & $0.39$\\
Fertility & (4.5K, 9.0K) & (21K, 124K) & $0.61$ & $0.29$\\
Fandisk & (6.5K, 13K) & (23K, 133K) & $4.75$ & $0.96$\\
Bunny & (7.1K, 14K) & (32K, 183K) & $4.17$ & $0.08$\\
Horse & (7.7K, 15K) & (34K, 198K) & $0.41$ & $0.35$\\
La Madeleine & (20K, 40K) & (68K, 421K) & $0.83$ & $0.24$\\
Beast & (30K, 61K) & (132K, 837K) & $0.57$ & $<0.01$\\
Armadillo & (50K, 100K) & (229K, 1.4M) & $0.63$ & $<0.01$\\
\midrule
Noisy-cube 5\% & (1.5K, 3.0K) & (29K, 139K) & $35.78$ & $<0.01$\\
Noisy-cube 10\% & (1.5K, 3.0K) & (30K, 149K) & $30.07$ & $0.05$\\
Noisy-cube 15\% & (1.5K, 3.0K) & (33K, 164K) & $21.08$ & $0.93$\\
Noisy-cube 20\% & (1.5K, 3.0K) & (33K, 167K) & $10.95$ & $3.73$\\
\bottomrule
\end{tabular}
}
\end{table}
\setlength{\tabcolsep}{6pt}

\subsection{Analysis of Context-Free Projection}
\label{sec:contextFreeProblems}

We implemented the four different context-free projection approaches in Figure~\ref{fig:historyFree}, and had 4 users informally test each, drawing a variety of curves on the various 3D models seen in this paper.
\rev{The pilots helped understand the limitations of context-free projections, as noted in Section~\ref{sec:contextFree} and illustrated in Figure~\ref{fig:contextFreeProblems}. Additional details about the pilot observations are given in \revm{Appendix}~\ref{app:contextFreePilots}.}

\rev{The most valuable insight was that the user stroke in mid-air often tended to {\bf mimic} the expected projected curve. Context-free approaches, by design, are unable to capture this mimicry, i.e., the notion that the change between projected point as we draw a stroke is commensurate with the change in the 3D points along the stroke. This observation motivated us to design projection methods that explicitly incorporate the shape of the mid-air stroke $\CP$ and the projected curve $\CQ$. We call these functions \emph{anchored}.}

%% file: 4_historical.tex
The limitations of context-free projection can be addressed by equipping stroke point projection with the context/history of recently drawn points and their projections. In this paper we minimally use only the most recent stroke point $\bp_{i-1}$ and its projection $\bq_{i-1}$, as context to {\bf anchor} the current projection.

\begin{figure}
    \centering
    \includegraphics[width=\linewidth]{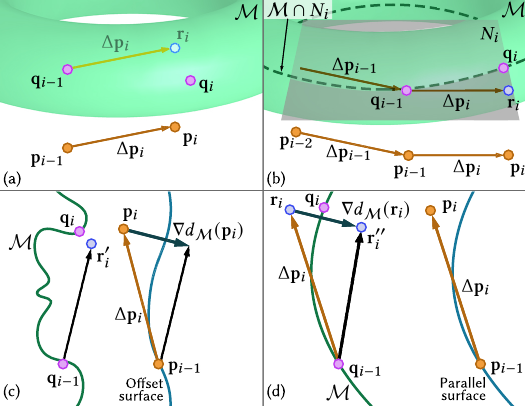}
    \caption{Anchored smooth-closest-point (a), and refinements: using a locally-fit plane (b), and anchor point constrained to an offset (c) or parallel surface (d). $\bq_i$, is obtained by projecting $\br_i$ (a), $\br_i'$ (c), or $\br_i''$ (d) onto $\CM$ via smooth-closest-point; or closest-point to $\br_i$ in $\CM\cap N_i$ (b). }
    \label{fig:historyProximity}
\end{figure}

Any reasonable context-free projection can be used for the first stroke point $\bp_0$. We use \emph{spraycan} $\pi_{spray}$, our preferred context-free technique. For subsequent points ($i>0$), we compute:
\begin{equation}
    \br_i = \bq_{i-1} + \Delta\bp_i,
\end{equation}

where $\Delta\bp_i=(\bp_i - \bp_{i-1})$. We then compute $\bq_i$ as a projection of the anchored stroke point $\br_i$ onto $\CM$, that attempts to capture $\Delta\bp_i \approx \Delta\bq_i$.
Anchored projection captures our observation that the mid-air user stroke tends to {\bf mimic} the shape of their intended curve on surface. 
While users to do not adhere consciously to any precise geometric formulation of mimicry, we observe that users often draw the intended projected curve as a corresponding stroke on an imagined offset or translated surface (Figure~\ref{fig:historyProximity}).
A good general projection for the anchored point $\br_i$ to $\CM$ thus needs to be continuous, predictable, and loosely capture this notion of mimicry.

\subsection{Mimicry Projection}
\label{sec:anchoredProximity}

Controller sampling rate in current \revm{VR} systems is 50Hz or more, meaning that even during ballistic movements, the distance $\|\Delta\bp_i\|$ for any stroke sample $i$ is of the order of a few millimetres.
Consequently, the anchored stroke point $\br_i$ is typically much closer to $\CM$, than the stroke point $\bp_i$, making closest-point \emph{snap} projection a compelling candidate for projecting $\br_i$.
Such an \emph{anchored closest-point} projection explicitly minimizes $\|\Delta\bp_i - \Delta\bq_i\|$, but precise minimization is less important than avoiding projection discontinuities and undesirably snapping, even for points close to the mesh. Our formulation of a \emph{smooth-closest-point} projection $\pi_{SCP}$ in \S~\ref{sec:phong} addresses these goals precisely.
We define \emph{mimicry} projection as

\begin{equation}
    \label{eq:acp}
    \Pi_{mimicry}(\bp_i) = 
    \begin{cases}
        \pi_{spray}(\bp_i) & \text{if } i=0, \\
        \pi_{SCP}(\br_i) & \text{otherwise.}
    \end{cases}
\end{equation}

\subsection{Refinements to Mimicry Projection}

We further explore refinements to \emph{mimicry} projection, that might improve curve projection in certain scenarios.

{\bf Planar curves} are very common in design and visualization \cite{mccrae2011slices}. We can locally encourage planarity in mimicry projection by constructing a plane $N_i$ with normal $\Delta\bp_i \times \Delta\bp_{i-1}$ (i.e. the local plane of the mid-air stroke) and passing through the anchor point $\br_i$ (Figure~\ref{fig:historyProximity}b).
We then intersect $N_i$ with $\CM$. $\bq_i$ is defined as the closest-point to $\br_i$ on the intersection curve that contains $\bq_{i-1}$. 
Note, we use $\pi_{spray}(\bp_i)$ for $i<2$, and we retain the most recently defined normal direction ($N_{i-1}$ or prior) when $N_i= \Delta\bp_i \times \Delta\bp_{i-1}$ is undefined.
We find this method works well for near-planar curves, but the plane is sensitive to noise in the mid-air stroke (Figure~\ref{fig:anchoredProblems}f), and can feel \emph{sticky} or less responsive for non-planar curves.

\textbf{Offset and parallel surface drawing} captures the observation that
 users tend to draw an intended curve as a corresponding stroke on an imaginary offset or parallel surface of the object $\CM$. While we do not expect users to draw precisely on such surfaces, \rev{it} is unlikely a user would intentionally draw orthogonal to \rev{them}.

In scenarios when a user is sub-consciously drawing on \rev{an} offset surface of $\CM$ (an isosurface of its signed-distance function $d_\CM(\cdot)$), we can remove the component of a user stroke segment along the gradient $\nabla d_\CM$, when computing the anchor point (Figure~\ref{fig:historyProximity}c):
\begin{equation}
    \label{eq:offsetSurface}
    \br_i' = \bq_{i-1} + \Delta\bp_i - \Big(\Delta\bp_i \cdot \nabla d_\CM(\bp_i) \Big)\nabla d_\CM(\bp_i)
\end{equation}
We can similarly locally constrain user strokes to a parallel surface of $\CM$ in Equation~\ref{eq:parallelSurface} as:
\begin{equation}
    \label{eq:parallelSurface}
    \br_i'' = \bq_{i-1} + \Delta\bp_i - \Big(\Delta\bp_i \cdot \nabla d_\CM(\br_i) \Big)\nabla d_\CM(\br_i)\text{.}
\end{equation}
Note that the difference from Eq.~\ref{eq:offsetSurface} is the position where $\nabla d_\CM$ is computed, as shown in Figure~\ref{fig:historyProximity}d.
A parallel surface better matched user expectation than an offset surface in our pilot testing, but both techniques produced poor results when user drawing deviated from these imaginary surfaces (Figure~\ref{fig:anchoredProblems}g--l).

\subsection{Anchored Raycast Projection}
\label{sec:anchoredRaycast}

\begin{figure}
    \centering
    \includegraphics[width=\linewidth]{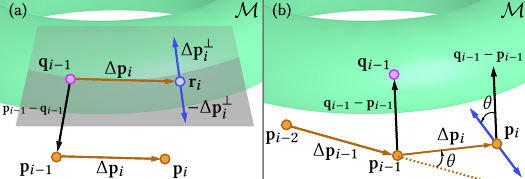}
    \caption{Anchored raycast techniques: ray direction defined orthogonal to $\Delta\bp_i$ in a local plane (a); parallel transport of ray direction along the user stroke (b). The cast rays (forward/backward) are shown in blue.}
    \label{fig:historyRaycast}
\end{figure}

For completeness, we also investigated raycast alternatives to projection of the anchored stroke point $\br_i$. We used similar priors of local planarity and offset or parallel surface transport as with mimicry refinement, to define ray directions. Figure~\ref{fig:historyRaycast} shows two such options.

In Figure~\ref{fig:historyRaycast}a, we cast a ray in the local plane of motion, orthogonal to the user stroke, given by $\Delta\bp_i$.
We construct the local plane containing $\br_i$ spanned by $\Delta\bp_i$ and $\bp_{i-1} - \bq_{i-1}$, and then define the direction orthogonal to $\Delta\bp_i$ in this plane. Since $\br_i$ may be inside $\CM$, we cast two rays bi-directionally $(\br_i, \pm \Delta\bp_i^\bot)$, where
\[\Delta\bp_i^\bot =
	\Delta\bp_i \times
	\big(
        \Delta\bp_i \times
        \left( \bp_{i-1} - \bq_{i-1} \right)
    \big)  
\]

If both rays successfully intersect $\CM$, we choose $\bq_i$ to be the point closer to $\br_i$.
As with locally planar mimicry projection, this technique suffered from instability in the local plane.

Motivated by mimicry, we also explored parallel transport of the projection ray direction along the user stroke \rev{(Figure~\ref{fig:historyRaycast}b)}.
For $i>0$, we parallel transport the previous projection direction $\bq_{i-1} - \bp_{i-1}$ along the mid-air curve by rotating it with the rotation that aligns $\Delta\bp_{i-1}$ with $\Delta\bp_{i}$. Once again\rev{,} bi-directional rays are cast from $\br_i$, and $\bq_i$ is set to the closer intersection with $\CM$. 

\begin{figure*}[htb]
    \centering
    \includegraphics[width=\linewidth]{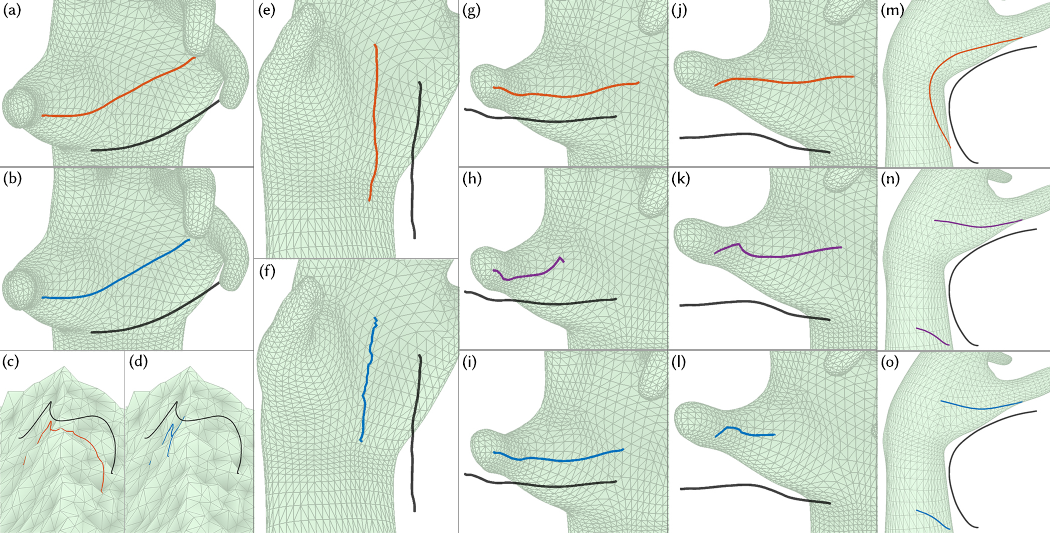}
    \caption{\emph{Mimicry} vs. other anchored stroke projections: Mid-air strokes are shown in \textbf{black} and \emph{mimicry} curves in \textcolor{matlabred}{\textbf{red}}. Anchored closest-point (\textcolor{matlabblue}{\textbf{blue}}), is similar to \emph{mimicry} on smooth, low-curvature meshes (a,b) but degrades with mesh detail/noise (c,d). Locally planar projection (\textcolor{matlabblue}{\textbf{blue}}) is susceptible local plane instability (e,f). Parallel (\textcolor{matlabpurple}{\textbf{purple}} h,k) or offset (\textcolor{matlabblue}{\textbf{blue}} i,l) surface based projection fail in (h,l) when the user stroke deviates from said surface, while \emph{mimicry} remains reasonable (g, j). Compared to \emph{mimicry} (m), anchored raycasting based on a local plane (\textcolor{matlabpurple}{\textbf{purple}} n), or ray transport (\textcolor{matlabblue}{\textbf{blue}} o) can be discontinuous.}
    \label{fig:anchoredProblems}
\end{figure*}

In general, we found that all raycast projections, even when anchored, suffered from 
unpredictability over long strokes and discontinuities when there are no ray-object intersections (Figure~\ref{fig:anchoredProblems}n,o).

\subsection{Final Analysis and Implementation Details}
\label{sec:implementation}

In summary, extensive pilot testing of the anchored techniques revealed that they \rev{were} generally better than context-free approaches, \rev{especially} when users drew further away from the 3D object. Among anchored techniques, stroke mimicry captured as an \emph{anchored-smooth-closest-point} projection proved to be theoretically elegant, and practically the most resilient to ambiguities of user intent and differences of drawing style among users.
\emph{Anchored closest-point} can be a reasonable proxy to \emph{anchored smooth-closest-point} when pre-processing is undesirable. \rev{\revm{A pertinent application is real-time sculpting,} where the object shape changes frequently.}

Our techniques are implemented in C\#, with interaction, rendering, and VR support provided by the Unity Engine.
For the smooth closest-point operation, we modified Panozzo et al.'s~\shortcite{panozzo2013weighted} reference implementation, which includes pre-processing code in MATLAB and C++, and real-time code in C++.
The real-time projection implementation is exposed to our C\# application via a compiled dynamic library.
\rev{In their implementation, as well as ours, $d=8$; that is, we embed $\CM$ in $\BR^8$.
Offset surfaces are computed using \texttt{libigl}~\cite{libigl}, with $\mu=\|\mathtt{BBox}(\CM)\|/20$.}
We then improve surface quality using TetWild~\cite{hu2018tetwild}, before computing the tetrahedral mesh $\CT_\CM$ using TetGen~\cite{si2015tetgen}.

We support fast closest-point queries, using an AABB tree implemented in \texttt{geometry3Sharp}~\cite{geometry3sharp}.
Signed-distance is also computed using the AABB tree and fast winding number~\cite{barill2018fast}, and gradient $\nabla d_\CM$ computed using central finite differences.

To ease replication of our various techniques and aid future work, \revm{we have released our open-source implementation at \href{https://github.com/rarora7777/curve-on-surface-drawing-vr}{\textcolor{blue}{\texttt{github.com/}}}
\href{https://github.com/rarora7777/curve-on-surface-drawing-vr}{\textcolor{blue}{\texttt{rarora7777/curve-on-surface-drawing-vr}}}}.

We now formally compare our most promising projection \emph{mimicry}, to the best state-of-the-art context-free projection \emph{spraycan}.

%% file: 5_study.tex
\begin{figure}
    \centering
    \includegraphics[width=\linewidth]{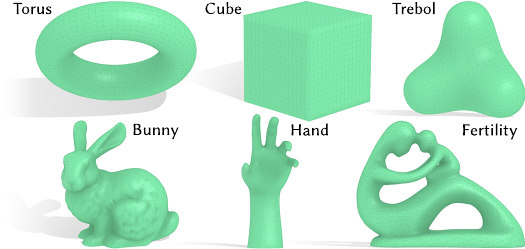}
    \caption{The six shapes utilized in the user study. The \emph{torus} shape was used for tutorials, while the rest were used for the recorded experimental tasks.}
    \label{fig:studyShapes}
\end{figure}

We designed a user study to compare the performance of the \emph{spraycan} and \emph{mimicry} methods for a variety of curve-drawing tasks.
We selected six shapes for the experiment (Figure~\ref{fig:studyShapes}), aiming to cover a diverse range of shape characteristics: sharp features (\emph{cube}), large smooth regions (\emph{trebol}, \emph{bunny}), small details with ridges and valleys (\emph{bunny}), thin features (\emph{hand}), and topological holes (\emph{torus}, \emph{fertility}).

\begin{figure}
    \centering
    \includegraphics[width=\linewidth]{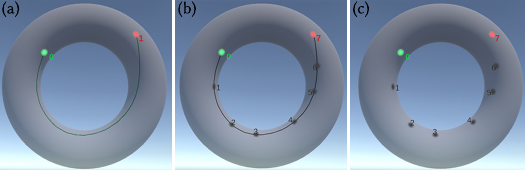}
    \caption{\rev{Study tasks---\emph{curve tracing}: target curve is visible when drawing (a), and \emph{curve re-creation}: target curve (b) is hidden when drawing (c).}}
    \label{fig:studyTasks}
\end{figure}

We then sampled ten distinct curves on the surface of each of the six objects.
A canonical task in our study involved the participant attempting to re-create a given \emph{target curve} from this set. We designed two types of drawing tasks
shown in Figure~\ref{fig:studyTasks}:\\
\textbf{Tracing curves}, where a participant tried to trace over a visible target curve  using a single smooth stroke.\\ 
\textbf{Re-creating curves}, where a participant attempted to re-create from memory, a visible target curve that was hidden as soon as the participant started to draw.  An enumerated set of keypoints on the curve\rev{,} however, remained as a visual reference, to aid the participant in re-creating the hidden curve with a single smooth stroke.

The rationale behind asking users to draw target curves is both to control the length, complexity, and nature of curves drawn by users, and to have an explicit 
representation of the user-intended curve. Curve tracing and re-creating are fundamentally different drawing tasks, each with important applications \cite{arora2017experimental}. Our curve re-creation task is designed to capture free-form drawing, with minimal visual suggestion of intended target curve.

\rev{Target curves were sampled randomly from a distribution of long, smooth, curves on the mesh. For each sample curve, 4--9 keypoints were selected along \revm{endpoints} and curvature extrema, the number depending on the curve's length and complexity. Positioning keypoints at curvature extrema ensured that curve re-creating tasks amounted to smoothly joining the keypoints, rather than testing participants' memory. Appendix~\ref{app:sampling} provides details about the sampling process.}

\subsection{Experiment Design}
\label{ref:expDesign}

The main variable studied in the experiment was \emph{Projection method}---\emph{spraycan} vs. \emph{mimicry}---realized as a within-subjects variable.
The order of methods was counterbalanced between participants.
For each method, participants were exposed to all the six objects.
Object order was fixed as torus, cube, trebol, bunny, hand, and fertility, based on our personal judgment of drawing difficulty.
The torus was used as a tutorial, where participants had access to additional

\noindent instructions visible in the scene and their strokes were not utilized for analysis.
For each object, the order of the 10 target strokes was randomized.
The first five were used for the tracing curves task, while the remaining five were used for re-creating curves.

The target curve for the first tracing task  was repeated after the five unique curves, to gauge user consistency and learning effects. A similar repetition 
was used for curve re-creation.
Participants thus performed 12 curve drawing tasks per object, leading to a total of $12 \times 5$ (objects) $\times\ 2$ (projections) $= 120$  strokes per participant.

Owing to the COVID-19 physical distancing guidelines, the study was conducted on participants' personal VR equipment at their homes. A 15-minute instruction video introduced the study tasks and the two projection methods.
Participants then filled out a consent form and a questionnaire to collect demographic information.
This was followed by them testing the first projection method and filling out a questionnaire to express their subjective opinions of the method.
They then tested the second method, followed by a similar questionnaire, and questions involving subjective comparisons between the two methods.
Participants were required to take a break after testing the first method, and were also encouraged to take breaks after drawing on the first three shapes for each method. The study took approximately an hour, including the questionnaires.

\subsection{Participants}
\label{participants}

Twenty participants (5 female\rev{, 15 male}) aged 21--47 from five countries participated in the study.
All but one were right-handed.
\revm{Participants were not selected for artistic ability or prior VR experience, and exhibited a diverse range of self-reported artistic abilities} (min. 1, max. 5, median 3 on a 1--5 scale) \revm{as well as} varying degrees of VR experience, ranging from below 1 year to over 5 years.
13 participants had a technical computer graphics or HCI background, while ten had experience with creative tools in VR, with one reporting professional usage.
Participants were paid $\approx22$ USD as a gift card.

\subsection{Apparatus}
\label{sec:apparatus}

As the study was conducted on personal VR setups, a variety of commercial VR devices were utilized---Oculus Rift, Rift S, and Quest using Link cable, HTC Vive and Vive Pro, Valve Index, and Samsung Odyssey using Windows Mixed Reality.
All but one participant used a standing setup allowing them to freely move around.

\subsection{Procedure}
\label{sec:procedure}

Before each trial, participants could use the ``grab'' button on their controller (in the dominant hand) to grab the mesh to position and orient it as desired.
The trial started as soon as the participant started to draw by pressing the ``main trigger'' on their dominant hand controller.
This action disabled the grabbing interaction---participants could not draw and move the object simultaneously.
As noted earlier, for curve re-creation tasks, this had the additional effect of hiding the target curve, but leaving keypoints visible.

%% file: 6_results.tex
We recorded the head position $\bh$ and orientation $\Ch$, controller position $\bc$ and orientation $\Cc$, projected point $\bq$, and timestamp $t$, for each mid-air stroke point $\bp = \bc$. 
\rev{We refer to a target curve as $\CX$, 
a mid-air stroke as $\CP$, and a projected curve as $\CQ$.}

\subsection{Data Processing and Filtering}
\label{sec:dataProcess}

We formulated three criteria to filter out meaningless user strokes.

\paragraph{Short Curves} We ignore projected curves $\CQ$ that are too short as compared to the length of the target curves $\CX$ (conservatively curves less than half as long as the target curve). While it is possible that the user stopped drawing mid-way out of frustration, we found it was more likely that they prematurely released the controller trigger by accident. 
Both curve lengths are computed in $\BR^3$ for efficiency.

\paragraph{Stroke Noise} We ignore strokes for which the mid-air stroke is too noisy.
Specifically, mid-air strokes with distant consecutive points 
($\exists\ i\ \text{s.t.}\ \|\bp_i - \bp_{i-1}\| > 5\text{cm}$) are rejected.

\paragraph{Inverted \rev{Curves}} While we labelled keypoints with numbers and marked start and end points in green and red (Figure~\ref{fig:studyTasks}), some users occasionally drew the target curve in reverse.
The motion to draw a curve in reverse is not symmetric, and such curves are thus rejected. We detect inverted strokes by \revm{looking} at the indices $i_0, i_1, \ldots, i_l$ of the points in $\CQ$ which are closest to the keypoints $\bx_{k_0}, \bx_{k_1}, \ldots, \bx_{k_l}$ of $\CX$. Ideally, the sequence $i_0,\ldots, i_l$ should have no inversions, i.e., $\forall\  0\le j<k\le l,\ i_j \le i_k$; and maximum $l(l+1)/2$ inversions, if $\CQ$ is aligned in reverse with $\CX$.
We consider curves with more than $l(l+1)/4$ (half the maximum) inversions to be inadvertently inverted and reject them.
\rev{Distances are computed in $\BR^3$ for efficiency.}

Despite conducting our experiment remotely without supervision, we found that \rev{95.8\%} of the strokes satisfied our criteria and could be utilized for analysis. \rev{Out of the 102 strokes deemed unfit for analysis, 17 were too short, 66 were inverted, and 38 exhibited excessive tracking noise.} \revm{It is possible that some of the short or inverted curves were caused due to curve control issues, there is no robust automatic method for distinguishing between inadvertent errors and genuine challenges faced by the users. Given the small number of such strokes and the potential bias in manual classification, we chose to exclude these strokes from the analysis.}
For comparisons between $\pi_{spray}$ and $\pi_{mimicry}$, we reject \rev{stroke} pairs where either \rev{stroke} did not satisfy the quality criteria.
Out of 1200 pairs (2400 total strokes), 1103 (91.9\%) satisfied the quality criteria and were used for analysis, including 564 pairs for the curve re-creation task and 539 for the tracing task.

\setlength{\tabcolsep}{3pt}
\begin{table}[tbh]
    \centering
    \caption{Quantitative results (mean $\pm$ std. dev.) of the comparisons between \emph{mimicry} and \emph{spraycan} projection.
    All measures are analyzed using Wilcoxon signed-rank tests, lower values are better, and significantly better values ($p<.05$) are shown in \textbf{boldface}.
    Accuracy, aesthetic, and physical effort measures are shown with \textcolor{tablegreen}{green},  \textcolor{tablered}{red}, and \textcolor{tableblue}{blue} backgrounds, respectively.
    }
    \label{tbl:quantitativeResults}
	\begin{tabular}{@{}lccrr@{}}
		\midrule
		\multicolumn{5}{@{}c@{}}{\large{\textsc{Tracing Curves}}}\\
		\midrule
		\textbf{Measure\hspace{-1ex}} & \textbf{Spraycan} & \textbf{Mimicry} & \textbf{$p$-value} & \textbf{$z$-stat} \\
		\rowcolor{tablegreen!30}
		$D_{ep}$ & 2.31 $\pm$ 2.64 mm & \textbf{1.13 $\pm$ 1.11 mm} & <.001 & 8.36 \\
		\rowcolor{tablegreen!15}
		$D_{sym}$ & 0.64 $\pm$ 0.66 mm & 0.56 $\pm$ 0.44 mm & >.05 & -0.09 \\
		\rowcolor{tablered!30}
		$K_E$ & 280 $\pm$ 262 rad/m & \textbf{174 $\pm$ 162 rad/m} & <.001 & 15.59 \\
		\rowcolor{tablered!15}
		$K_g$ & 249 $\pm$ 245 rad/m & \textbf{152 $\pm$ 157 rad/m} & <.001 & 15.42 \\
		\rowcolor{tablered!30}
		$F_g$ & 394 $\pm$ 413 rad/m & \textbf{248 $\pm$ 285 rad/m} & <.001 & 14.82 \\
		\rowcolor{tableblue!15}
		$T_h$ & 0.81 $\pm$ 0.70  & \textbf{0.58 $\pm$ 0.40 } & <.001 & 7.93 \\
		\rowcolor{tableblue!30}
		$R_h$ & 1.63 $\pm$ 2.18 rad/m & \textbf{1.18 $\pm$ 1.63 rad/m} & <.001 & 4.82 \\
		\rowcolor{tableblue!15}
		$T_c$ & \textbf{1.05 $\pm$ 0.36 } & 1.10 $\pm$ 0.29  & <.001 & -3.36 \\
		\rowcolor{tableblue!30}
		$R_c$ & 5.12 $\pm$ 5.88 rad/m & \textbf{3.79 $\pm$ 4.84 rad/m} & <.001 & 5.51 \\
		\rowcolor{tableblue!15}
		$\tau$ & \textbf{4.69 $\pm$ 1.85 s} & 5.29 $\pm$ 2.17 s & <.001 & -7.32 \\
		\midrule
		\multicolumn{5}{@{}c@{}}{\large{\textsc{\rev{Re-creating Curves}}}}\\
		\midrule
		\textbf{Measure\hspace{-1ex}} & \textbf{Spraycan} & \textbf{Mimicry} & \textbf{$p$-value} & \textbf{$z$-stat} \\
		\rowcolor{tablegreen!30}
		$D_{ep}$ & 2.34 $\pm$ 2.49 mm & \textbf{2.24 $\pm$ 23.32 mm} & <.001 & 8.63 \\
		\rowcolor{tablegreen!15}
		$D_{sym}$ & 0.75 $\pm$ 0.65 mm & 1.12 $\pm$ 11.51 mm & >.05 & 0.55 \\
		\rowcolor{tablered!30}
		$K_E$ & 254 $\pm$ 236 rad/m & \textbf{155 $\pm$ 127 rad/m} & <.001 & 14.70 \\
		\rowcolor{tablered!15}
		$K_g$ & 223 $\pm$ 219 rad/m & \textbf{132 $\pm$ 123 rad/m} & <.001 & 14.95 \\
		\rowcolor{tablered!30}
		$F_g$ & 348 $\pm$ 371 rad/m & \textbf{215 $\pm$ 227 rad/m} & <.001 & 14.11 \\
		\rowcolor{tableblue!15}
		$T_h$ & 0.72 $\pm$ 0.54  & \textbf{0.54 $\pm$ 0.35 } & <.001 & 6.78 \\
		\rowcolor{tableblue!30}
		$R_h$ & 1.50 $\pm$ 2.19 rad/m & \textbf{1.32 $\pm$ 1.99 rad/m} & .002 & 3.07 \\
		\rowcolor{tableblue!15}
		$T_c$ & \textbf{1.05 $\pm$ 0.37 } & 1.11 $\pm$ 0.23  & <.001 & -5.94 \\
		\rowcolor{tableblue!30}
		$R_c$ & 5.23 $\pm$ 6.36 rad/m & \textbf{3.63 $\pm$ 5.13 rad/m} & <.001 & 4.00 \\
		\rowcolor{tableblue!15}
		$\tau$ & \textbf{4.33 $\pm$ 1.57 s} & 4.92 $\pm$ 1.89 s & <.001 & -7.12 \\
	\end{tabular}
\end{table}
\setlength{\tabcolsep}{6pt}

\subsection{Quantitative Analysis} 
\label{sec:quantitative}

We define 10 different statistical measures (Table~\ref{tbl:quantitativeResults}) to compare $\pi_{spray}$ and $\pi_{mimicry}$ curves in terms of their accuracy, aesthetic, and effort in curve creation. 
We consistently use the non-parametric Wilcoxon signed rank test for all quantitative measures instead of a parametric test such as the paired $t$-test,
since the recorded data for none of our measures was normally distributed (normality hypothesis rejected via the Kolmogorov-Smirnov test, $p<.005$).
\rev{In addition, we analyze users' tendency to mimic the target strokes and consistency between repeated strokes in Appendix~\ref{app:quantitative}.}

\subsubsection{Curve Accuracy}
\label{sec:strokeAccuracy}
Accuracy is computed using two measures of distance between points on the projected curve $\CQ$ and target curve $\CX$. Both curves are densely re-sampled using $m=101$ sample points equi-spaced by arc-length.

Given $\CQ = {\bq_0, \ldots, \bq_{m-1}}$ and $\CX = {\bx_0, \ldots, \bx_{m-1}}$,
we compute the \emph{average equi-parameter distance} $D_{ep}$ as
\begin{equation}
    \label{eq:distanceEquiparameter}
    D_{ep}(\CQ) = \frac1m \sum_{i=0}^{m-1} d_E\left( \bq_i, \bx_i \right) \text{,}
\end{equation}
where $d_E$ computes the Euclidean distance between two points in $\BR^3$.
We also compute the \emph{average symmetric distance} $D_{sym}$ as
\begin{equation*}
    \label{eq:distanceSymmetric}
    D_{sym}(\CQ) = \frac1{2m} \sum_{i=0}^{m-1}\left( \min_{\bx\in X} d_E\left( \bq_i, \bx \right) \right) + 
     \frac1{2m} \sum_{i=0}^{m-1}\left( \min_{\bq\in Q} d_E\left( \bq, \bx_i \right) \right)
\end{equation*}
In other words, $D_{ep}$ computes the distance between corresponding points on the two curves and $D_{sym}$ computes the average minimum distance from each point on one curve to the other curve.



\begin{figure}
    \centering
    \begin{subfigure}{\linewidth}
        \centering
        \includegraphics[width=0.44\linewidth]{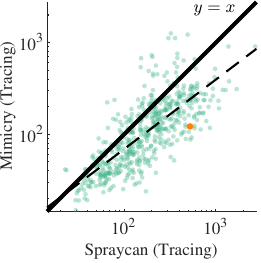}
        \includegraphics[width=0.44\linewidth]{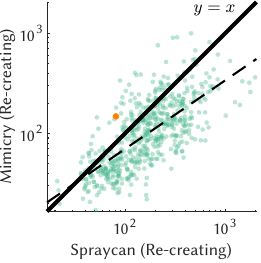}
        \vspace{-0.2cm}
        \caption{Normalized geodesic curvature $K_g$.}
		\vspace{0.2cm}
    \end{subfigure}
    \begin{subfigure}{\linewidth}
        \centering
        \includegraphics[width=0.44\linewidth]{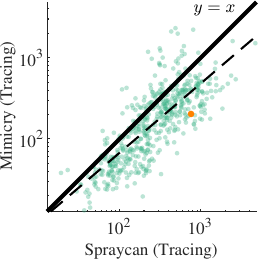}
        \includegraphics[width=0.44\linewidth]{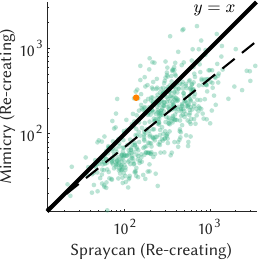}
        \vspace{-0.2cm}
        \caption{Normalized fairness deficiency $F_g$.}
		\vspace{0.2cm}
    \end{subfigure}
    \begin{subfigure}{\linewidth}
        \centering
        \includegraphics[width=0.88\linewidth]{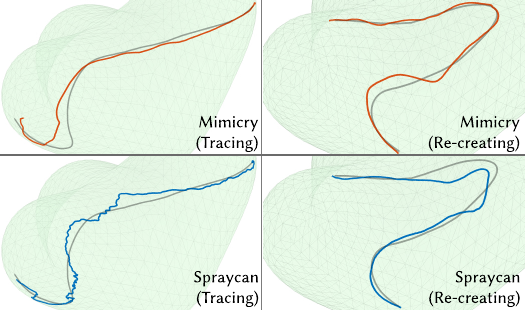}
		\vspace{-0.1cm}
        \caption{Example strokes, \textcolor{matlaborange}{\textbf{orange}} points in (a, b) above.}
    \end{subfigure}
    
    \caption{Curvature measures (a,b) indicate that \emph{mimicry} produces  significantly smoother and fairer curves than \emph{spraycan} for both tracing (left) and re-creating tasks (right). Pairwise comparison plots between \emph{mimicry} (y-axis) and \emph{spraycan} (x-axis), favour \emph{mimicry} for the vast majority of points (points below the $y=x$ line). A linear regression fit (on the log plots) is shown as a dashed line.
	Example curve pairs (\textcolor{matlaborange}{\textbf{orange}} points) for curve tracing and re-creating are also shown with the target curve $\CX$ shown in gray (c).}
    \label{fig:smoothnessResult}
\end{figure}

For both tracing and re-creation tasks, $D_{ep}$ indicated that \emph{mimicry} produced significantly better results than \emph{spraycan} (see Table~\ref{tbl:quantitativeResults}, Figure~\ref{fig:teaser}c,~\ref{fig:smoothnessResult}). 
The $D_{sym}$ difference was not statistically significant, evidenced by users correcting their strokes to stay close to the intended target curve (at the expense of curve aesthetic).

\subsubsection{Curve Aesthetic}
\label{sec:strokeQuality}

For most design applications, jagged projected curves, even if geometrically quite accurate, are aesthetically undesirable~\cite{mccrae2008sketching}.
Curvature-based measures are typically used to measure fairness of curves. We report three such measures of curve aesthetic for the projected curve $\CQ=\bq_0,\ldots, \bq_{n-1}$.
We first refine $\CQ$ by computing the exact geodesic on $\CM$ between consecutive points of $\CQ$ ~\cite{surazhsky2005fast}, to create $\HCQ$ with points $\Hbq_0, \ldots, \Hbq_{k-1}$, $k\ge n$. We choose to normalize our curvature measures using $L_{\CX}$, the length of the corresponding target stroke $\CX$.
The \emph{normalized Euclidean curvature} for $\CQ$ is defined as
\begin{equation}
    \label{eq:euclideanCurvature}
    K_E(\CQ) = \frac1{L_{\CX}} \sum_{i=1}^{\rev{k-2}} \theta_i
\end{equation}
where $\theta_i$ is the angle between the two segments of $\HCQ$ incident on $\Hbq_i$.
Thus, $K_E$ is the total discrete curvature of $\HCQ$, normalized by the target curve length.

Since $\HCQ$ is embedded in $\CM$, we can also compute discrete \emph{geodesic} curvature, computed as the deviation from the straightest geodesic on a surface. Using a signed $\theta^g_i$ defined at each point $\Hbq_i$ \cite{polthier2006straightest}, we compute \emph{normalized geodesic curvature} as 
\begin{equation}
    \label{eq:geodesicCurvature}
    K_g(\CQ) = \frac1{L_{\CX}} \sum_{i=1}^{\rev{k-2}} |\theta_i^g|\text{.}
\end{equation}

Finally, we define fairness~\cite{arora2017experimental, mccrae2008sketching} as a first-order variation in geodesic curvature, thus defining the \emph{normalized fairness deficiency} as

\begin{equation}
    \label{eq:fairness}
    F_g(\CQ) = \frac1{L_{\CX}} \sum_{i=2}^{\rev{k-2}} |\theta_i^g - \theta_{i-1}^g|\text{,}
\end{equation}

For all three measures, a lower value indicates a smoother, pleasing, curve.
Wilcoxon signed-rank tests on all three measures indicated that \emph{mimicry} produced significantly smoother and better curves than \emph{spraycan} (Table~\ref{tbl:quantitativeResults}).

\subsubsection{Physical Effort}
\label{sec:effortQuantitative}
The amount of head (HMD) and hand (controller) movement, and stroke \emph{execution time} $\tau$ \rev{provide quantitative} proxies for physical effort.

For head and hand translation, we first filter the position data with a Gaussian-weighted moving average filter with $\sigma=20\text{ms}$.
We then define \emph{normalized head/controller translation} $T_h$ and $T_c$ as the length of the poly-line defined by the filtered head/controller positions normalized by the length of the target curve $L_\CX$.

An important ergonomic measure is the amount of head/hand rotation required to draw the mid-air stroke.
We first de-noise or filter the forward and up vectors of the head/controller frame, using the same filter as for positional data.
We then re-orthogonalize the frames and compute the length of the curve defined by the filtered orientations in $\mathrm{SO(3)}$, using the angle between consecutive orientation data-points. 
We define \emph{normalized head/controller rotation} $R_h$ and $R_c$ as its orientation curve length, normalized by $L_\CX$.


Table~\ref{tbl:quantitativeResults} summarizes the physical effort measures.
We observe lower controller translation (effect size $\approx 5\%$) and execution time (effect size $\approx 12\%$) in favour of \emph{spraycan}; lower head translation and orientation (effect sizes $\approx 36\%, 26\%$) in favour of \emph{mimicry}. Noteworthy is the significantly reduced controller rotation using \emph{mimicry}, with \emph{spraycan} unsurprisingly requiring $35\%$ (tracing) and 44\% (re-creating) more hand rotation from the user.

\begin{figure}
        \centering
        \includegraphics[width=\linewidth]{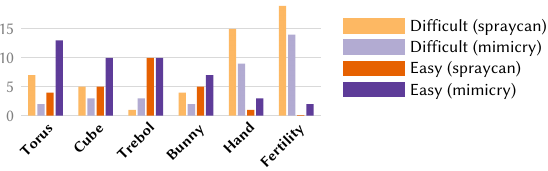}
    \caption{Perceived difficulty of drawing for the six 3D shapes in the study.}
    \label{fig:difficulty}
\end{figure}

\begin{figure}
    \centering
    \begin{subfigure}{\linewidth}
        \centering
        \includegraphics[width=\linewidth]{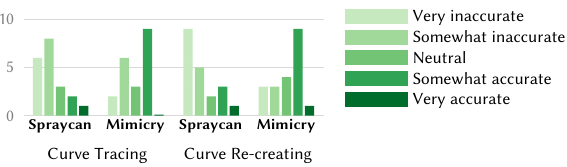}
		\vspace{-.5cm}
        \caption{Perceived accuracy.}
    \end{subfigure}
    \begin{subfigure}{\linewidth}
        \centering
        \includegraphics[width=\linewidth]{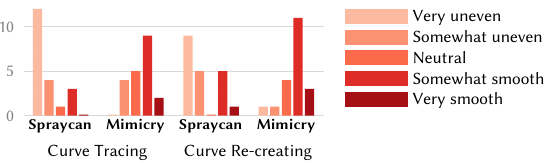}
		\vspace{-.5cm}
        \caption{Perceived smoothness.}
    \end{subfigure}
        \begin{subfigure}{\linewidth}
        \centering
        \includegraphics[width=\linewidth]{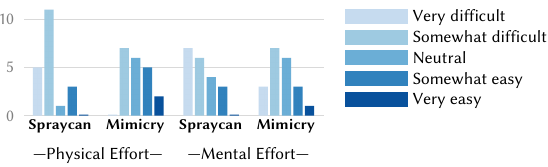}
		\vspace{-.5cm}
        \caption{Physical and mental effort ratings.}
    \end{subfigure}
    \caption{Participants perceived \emph{mimicry} to be better than \emph{spraycan} in terms of accuracy (a), curve aesthetic (b) and user effort (c).}
    \label{fig:qualitativeComparison}
\end{figure}

\begin{figure}
    \centering
    \includegraphics[width=\linewidth]{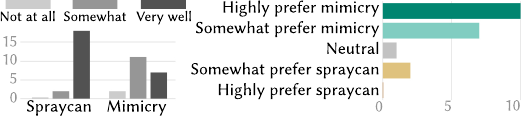}
    \caption{Participants stated understanding \emph{spraycan} projection better (left); 17/20 users stated an overall preference for \emph{mimicry} over \emph{spraycan} (right).}
    \label{fig:prefenceUnderstanding}
\end{figure}

\begin{figure*}
    \centering
    \includegraphics[width=\linewidth]{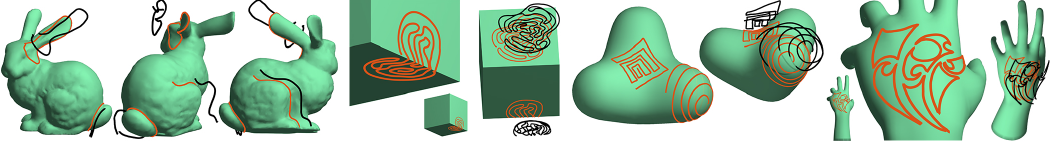}
    \caption{Gallery of free-form curves in \textcolor{matlabred}{\textbf{red}}, drawn \rev{by the paper authors} using \emph{mimicry}. (Left to right) tracing geometric features on the bunny, maze-like curves on the cube, maze with sharp corners and a spiral on the trebol, and artistic tattoo motifs on the hand. Some mid-air strokes (\textbf{black}) hidden for clarity.}
    \label{fig:curveGallery}
\end{figure*}

\subsection{Qualitative Analysis}
\label{sec:qualitative}


The mid- and post-study questionnaires elicited qualitative responses from participants on their perceived difficulty of drawing, curve accuracy and smoothness, mental and physical effort, understanding of the projection methods, and overall method of preference.

\rev{Participants were asked to specify the objects which they found especially easy or difficult to draw on, when using either of the two projection methods. In general, the shapes shown earlier were judged to be easier to work with (Figure~\ref{fig:difficulty}), validating our ordering of shapes in the experiment based on expected drawing difficulty. Importantly, this observation also suggests a lack of any learning effects caused by the fixed object ordering.}

Accuracy, smoothness, physical/mental effort responses were collected via 5-point Likert scales. We consistently order the choices from 1 (worst) to 5 (best) in terms of user experience, and report median ($M$) scores here.
\emph{Mimicry} was perceived to be a more accurate projection (tracing, re-creating $M=3,3.5$) compared to \emph{spraycan} ($M=2,2$), with 9 participants perceiving their traced curves to be either \emph{very accurate} or \emph{somewhat accurate} with \emph{mimicry}, compared to 2 for \emph{spraycan} (Figure~\ref{fig:qualitativeComparison}a).
Perception of stroke smoothness was also consistent with quantitative results, with \emph{mimicry} (tracing, re-creating $M=4,4$) clearly outperforming \emph{spraycan} (tracing, re-creating $M=1,2$) (Figure~\ref{fig:qualitativeComparison}b).
Lastly, with no need for controller rotation, \emph{mimicry} ($M=3$) was perceived as less physically demanding than \emph{spraycan} ($M=2$), as expected (Figure~\ref{fig:qualitativeComparison}c).

The response to understanding and mental effort was more complex. \emph{Spraycan}, with its physical analogy and mathematically precise definition was clearly understood by all 20 participants (17 very well, 3 somewhat)  
(Figure~\ref{fig:prefenceUnderstanding}a). \emph{Mimicry}, conveyed as ``drawing a mid-air stroke on or near the object as similar in shape as possible to the intended projection'', was less clear to users (7 very well, 11 somewhat, 3 not at all). Despite not understanding the method \rev{consciously}, the 3 participants were able to create curves that were both accurate and smooth. Further, users perceived \emph{mimicry} ($M=2.5$) as less cognitively taxing than \emph{spraycan} ($M=2$) (Figure~\ref{fig:qualitativeComparison}c). We believe this may be because users were less prone to consciously controlling their stroke direction and rather focused on drawing. The tendency to mimic may have thus manifested sub-consciously, as we had observed in pilot testing.
 
The most important qualitative question was user preference (Figure~\ref{fig:prefenceUnderstanding}b).
$85\%$ of the 20 participants preferred \emph{mimicry} (10 highly preferred, 7 somewhat preferred). The remaining users were neutral (1/20) or somewhat preferred \emph{spraycan} (2/20).

\subsection{Participant Feedback}
We also asked participants to elaborate on their stated preferences and ratings.
Participants (\emph{P4,8,16,17}) noted discontinuous \pquote{jumps} caused by \emph{spraycan}, and felt the continuity guarantee of \emph{mimicry}: \pquote[P6]{seemed to deal with the types of jitter and inaccuracy VR setups are prone to better}; \pquote[P9]{could stabilize my drawing}.
\emph{P9,15} felt that \emph{mimicry} projection was smoothing their strokes (no smoothing was employed): we believe this may be the effect of noise and inadvertent controller rotation, which \emph{mimicry} ignores, but can cause 
large variations with \emph{spraycan}, perceived as curve smoothing.

Some participants (\emph{P4,17}) felt that rotating the hand smoothly while drawing was difficult, while others missed the \emph{spraycan} ability to simply use hand rotation to sweep out long projected curves from a distance (\emph{P2,7}).
Participants commented on physical effort: \pquote[P4]{Mimicry method seemed to required [sic] much less head movement, hand rotation and mental planning}.

Participants appreciated the anchored control of \emph{mimicry} in high-curvature regions (\emph{P1,2,4,8}) also noting that with \emph{spraycan}, \pquote[P1]{the curvature of the surface could completely mess up my stroke}.
Some participants did feel that \emph{spraycan} could be preferable when drawing on near-flat regions of the mesh (\emph{P3,14,19,20}).

Finally, participants who preferred spraycan felt that mimicry required more thinking: \pquote[P3]{with mimicry, there was extra mental effort needed to predict where the line would go on each movement}, or because mimicry felt \pquote[P7]{unintuitive} due to their prior experience using a \emph{spraycan} technique. Some who preferred \emph{mimicry} found it difficult to use initially, but felt it got easier over the course of the experiment (\emph{P4,17}).

%% file: 7_applications.tex
\begin{figure}
	\centering
	\includegraphics[width=\linewidth]{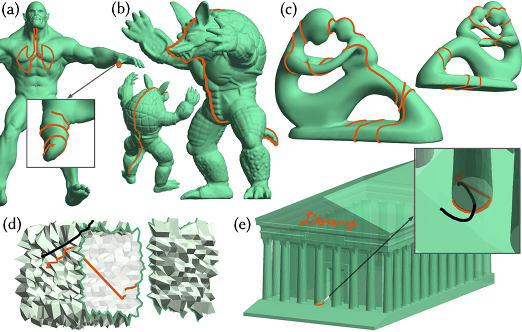}
	\caption{\rev{\emph{Mimicry} can be used to draw long, complicated curves on complex high-resolution meshes. We show strokes on high-resolution meshes (a, b, e), a long stroke bisecting a model (b), and a single stroke winding around a topologically non-trivial object multiple times (c). However, excessive noise in the input mesh can break the underlying $\pi_{ACP}$ assumptions, resulting in catastrophic failure (d). Mesh has been cut open for visualization. \revm{Large meshes with many sharp features and topological complexity can also show smaller local failures in the form of unexpected jumps when drawing close to the sharp features (e, inset).}}}
	\label{fig:stressTests}
\end{figure}

\begin{figure*}
    \centering
    \includegraphics[width=\linewidth]{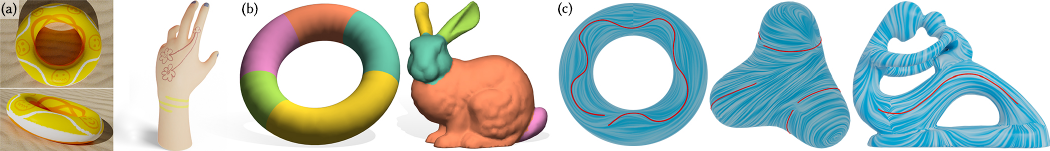}
    \caption{\rev{Applications of \emph{mimicry} projection. Texture painting (a), interactive segmentation by drawing curves onto meshes (b), and providing constraints (\textcolor{matlabred}{\textbf{red}} curves) to guide the vector field generation of Fisher et al.~\shortcite{fisher2007design} (c).}}
    \label{fig:applications}
\end{figure*}

Complex 3D curves on arbitrary surfaces can be drawn in \revm{VR} with a single stroke, using \emph{mimicry} (Figure~\ref{fig:curveGallery}).
Drawing such curves on 3D virtual objects is fundamental to many applications, including direct painting of textures ~\cite{schmidt2006discrete}; tangent vector field design~\cite{fisher2007design}; texture synthesis~\cite{turk2001texture, lefebre2006appearance}; interactive selection, annotation, and object segmentation~\cite{chen2009benchmark}; and seams for shape parametrization~\cite{levy2002least, rabinovich2017scalable, sawhney2017boundary}, registration~\cite{gehre2018interactive}, and quad meshing~\cite{tong2006designing}.
We showcase the utility and quality of \emph{mimicry} curves within example applications (also see supplemental video).

\rev{We also stress-test our technique by drawing curves on complex models (Figure~\ref{fig:stressTests}a,b,e) and drawing a single long curve looping around the fertility model multiple times (Figure~\ref{fig:stressTests}c). Finally, we show a failure case discussed in \S~\ref{sec:phong}---the \emph{mimicry} projection fails catastrophically due to problems in the underlying $\pi_{ACP}$ projection when the mesh is perturbed with excessive random noise \revm{(Figure~\ref{fig:stressTests}d). Smaller local jumps can also occur when the model is both highly detailed and contains many sharp features (see inset in Figure~\ref{fig:stressTests}e).}}

\paragraph{Texture Painting} Figures~\ref{fig:teaser}e, \ref{fig:applications}a show examples of textures painted in VR using \emph{mimicry}.
The long, smooth, wraparound curves on the torus, are especially hard to draw with 2D interfaces.
Our implementation uses Discrete Exponential Maps (DEM)~\cite{schmidt2006discrete} to compute a dynamic local parametrization around each projected point $\bq_i$, to create brush strokes or geometric stamps on the object.

\paragraph{Mesh Segmentation}
\rev{Figures~\ref{fig:teaser}e, \ref{fig:applications}b show interactive segmentation using \emph{mimicry}.}
In our implementation\rev{,} users draw an almost-closed curve $\CQ = \{\bq_0,\ldots,\bq_{n-1}\}$ on the object using \revm{\emph{mimicry}}. We snap points $\bq_i$ to their nearest mesh vertex, and use Dijkstra's shortest path to connect consecutive vertices, and to close the cycle. \rev{While easy in \revm{VR} via \emph{mimicry}, drawing similar strokes in 2D for selection/segmentation would require multiple view changes.}

\paragraph{Vector Field Design}
Vector fields on meshes are commonly used for 
texture synthesis~\cite{turk2001texture}, guiding fluid simulations~\cite{stam2003flows}, and non-photorealistic rendering~\cite{hertzmann2003illustrating}.
We use \emph{mimicry} curves as soft constraints to guide the vector field generation of Fisher et al.~\shortcite{fisher2007design}.
Figure~\ref{fig:applications}c shows example vector fields, visualized using Line Integral Convolutions~\cite{cabral1993imaging} in the texture domain.

%% file: 8_conclude.tex

We have presented a detailed investigation of the problem of real-time inked drawing on 3D virtual objects in immersive environments. We show the importance of stroke context when projecting mid-air 3D strokes, and explore the design space of anchored projections. A 20-participant study showed \emph{mimicry} to be preferred over the established \emph{spraycan} projection for projecting 3D strokes onto objects in \revm{VR}. Both \emph{mimicry} projection and performing VR studies in the wild do have some limitations. Further, while user stroke processing for 2D interfaces is a mature field of research, mid-air stroke processing for AR/VR is relatively nascent, with many directions for future work. \rev{Our study contributes a high-quality VR data corpus comprising $\approx 2400$ user strokes, projected curves, intended target curves, and corresponding system states, useful for future data-driven techniques for mid-air stroke processing.}

\paragraph{``In the wild'' VR Study Limitations}
Ongoing pandemic restrictions presented both a challenge and an opportunity to remotely conduct a more natural study in the wild, with a variety of consumer VR hardware and setups. The enthusiasm of the VR community allowed us to readily recruit 20 diligent users, albeit with a bias towards young, adult males.
\rev{While the variation in VR headsets seemed to be of little consequence, differences in controller grip and weight can certainly impact mid-air drawing posture and stroke behavior.} Controller size is also significant: a larger Vive controller, for example, has a higher chance of occluding target objects and projected \revm{curves}, as compared to a smaller Oculus Touch controller.
We could have mitigated the impact of controller size by rendering a standard drawing tool, but we preferred to render the familiar, default controller that matched the physical device in participants' hands. Further, no participant explicitly mentioned the controller getting in the way of their ability to draw.


\paragraph{Mimicry Limitations}
Our lack of a concise mathematical definition of observed stroke mimicry, makes it harder to precisely communicate it to users. While a precise mathematical formulation may exist, conveying it to non-technical users can still be a challenging task. 
\emph{Mimicry} ignores controller orientation, producing smoother strokes with less effort, but can give participants a reduced sense of sketch control (\emph{P2,3,6}). We hypothesize that the reduced sense of control is in part due to the tendency for anchored smooth-closest-point to shorten the user stroke upon projection, sometimes creating a feeling of lag. \emph{Spraycan} like techniques, in contrast, have a sense of amplified immediacy, and the explicit ability to make lagging curves catch-up by rotating a controller in place.

\paragraph{Future work}
Our goal was to develop a general real-time inked projection with minimal stroke context via anchoring.
Optimizing the method to account for the entire partially projected stroke may improve the projection quality. Relaxing the restriction of real-time inking would allow techniques such as spline fitting and global optimization that can account for the entire user stroke and geometric features of the target object. Local parametrizations such as DEM (\S~\ref{sec:applications}) can be used to incrementally grow or shrink the projected curve, so it does not lag the user stroke. Hybrid projections leveraging both proximity and raycasting are also subject to future work.

On the interactive side, we \revm{experimented} with feedback to encourage users to draw closer to a 3D object. For example, we tried varying the appearance of the line connecting the controller to the projected point based on line length; or providing aural/haptic feedback if the controller got further than a certain distance from the object. While these techniques can help users in specific drawing or tracing tasks, we found them to be distracting and harmful to stroke quality for general stroke projection. Bimanual interaction, such as rotating the shape with one hand while drawing on it with the other (suggested by \emph{P3,19}), can also be explored. \rev{\revm{To generalize our work to AR}, the impact of rendering quality and perception of virtual models also needs to be studied in the future. Drawing on physical objects in AR is another related research direction.}

\rev{Application-dependent optimizations to encourage closed strokes, snapping to geometric features, or alignment with existing user-drawn curves, can also be explored in the future.} \revm{Further, our user study only focused on smooth curves. While we show author-drawn example of curves with sharp features (Figure~\ref{fig:curveGallery}), formally testing the \emph{mimicry} technique for drawing such curves and potentially optimizing the projection to deal with sharp features is an important future direction.} But perhaps the most exciting area of future work is data-driven techniques \rev{for inferring} the intended projection, perhaps customized to the drawing style of individual users. Our \revm{study code and data has been made publicly available at \href{https://github.com/rarora7777/curve-on-surface-drawing-vr}{\textcolor{blue}{\texttt{github.com/rarora7777/curve-on-surface-drawing-vr}}}} to aid in such endeavours.


\begin{acks}

We are thankful to Michelle Lei for developing the initial implementation of the context-free techniques, and to Jiannan Li and Debanjana Kundu for helping pilot our methods.
We also thank various 3D model creators and repositories for the models we utilized: Stanford bunny and armadillo models courtesy of the Stanford 3D Scanning Repository, trebol model provided by Shao et al.~\shortcite{shao2012crossshade}, fertility model courtesy the Aim@Shape repository, hand model provided by \texttt{Jeffo89} on \url{turbosquid.com}, horse model courtesy Cyberware, Spiderman bust base model by David Ruiz Olivares (CC BY 4.0), beast model courtesy Autodesk, fandisk model provided by Pratt \& Whitney/Hughes Hoppe, La Madeleine model by LeFabShop on \url{thingiverse.com} (CC BY-SA 3.0), and cup model (Figure~\ref{fig:2DProblems}) provided by Daniel Noree on \url{thingiverse.com} (CC BY 4.0).

This work has been funded by \grantsponsor{nserc}{NSERC}{https://www.nserc-crsng.gc.ca/index_eng.asp} \grantnum{nserc}{Discovery Grant 480538}, and by software and research donations from Adobe.

\end{acks}

%% file: 99_appendix.tex
\appendix

\section{Context-Free Pilot Observations}
\label{app:contextFreePilots}

In this appendix, we provide additional informal observations from our pilot tests with context-free techniques (Section~\ref{sec:contextFree}), as well as additional details on the limitations of such techniques.

\subsection{Qualitative observations}
\begin{itemize}[label=--, leftmargin=1em]
\item \emph{Head-centric} and \emph{occlude} projections become unpredictable if the user is inadvertently changing their viewpoint while drawing. These projections are also only effective when drawing frontally on an object, like with a 2D interface. Neither as a result exploits the potential gains of mid-air drawing in AR/VR.
\item \emph{Spraycan} projection was clearly the most effective context-free technique.
We noted however, that consciously reorienting the controller while drawing on or around complex objects was both cognitively and physically tiring.
\item \emph{Snap} projection was quite sensitive to changes in the distance of the stroke from the object surface, and in general produced the most undulating projections due to closest-point singularities.
\item All projections converge to the mid-air user stroke when it precisely conforms to the surface of the 3D object. But as the distance between the object and points on the mid-air stroke increases, their behavior diverges quickly.
\item While users did draw in the vicinity and mostly above the object surface, they rarely drew precisely on the object. The average distance of stroke points from the target object was observed to be 4.8 cm in a subsequent user study (\S~\ref{sec:study}).
\end{itemize}

\subsection{Details on the Limitations of Context-Free Methods}
\label{app:contextFreeLimitations}

The inability of context-free approaches to capture a notion of stroke mimicry---due to a lack of curve history or context---materializes as problems in different forms.

\subsubsection{Projection Discontinuities}
\label{app:discontinuity}
Proximal projection (including \emph{smooth-closest-point}) can be highly discontinuous with increasing distance from the 3D object, particularly in concave regions (Figure~\ref{fig:contextFreeProblems}a).
Mid-air drawing along valleys without staying in precise contact with virtual object is thus
extremely difficult.
Raycast projections can similarly suffer large discontinuous jumps across occluded regions (in the ray direction) of the object (Figure~\ref{fig:contextFreeProblems}d).

While this problem theoretically exists in 2D interfaces as well, it is less observed in practice for two reasons: 2D drawing on a constraining physical surface is significantly more precise than mid-air drawing in AR/VR~\cite{arora2017experimental}; and artists minimize such discontinuities by carefully choosing appropriate views (raycast directions) before drawing each curve. Automatic direction control of view or controller, while effective in 2D \cite{ortega2014direct}), is detrimental to a sense of agency and presence in AR/VR.

\subsubsection{Undesirable Snapping}
\label{app:undesirableSnapping}
Proximity-based methods also tend to get stuck on sharp (or high curvature) convex features of the object (Figure~\ref{fig:contextFreeProblems}b).
While this can be useful to trace along a ridge feature, it is particularly problematic for general curve-on-surface drawing.

\subsubsection{Projection depth disparity}
\label{app:geometryIgnorance}
The relative orientation between the 3D object surface and raycast direction can cause large depth disparities between parts of user strokes and curves projected by raycasting (Figure~\ref{fig:contextFreeProblems}c). Such irregular bunching or spreading of points on the projected curve also goes against our observation of stroke mimicry. Users can arguably reduce this disparity by continually orienting the view/controller to keep the projection ray well aligned with object surface normal. Such re-orientation however can be tiring, ergonomically awkward, and deviates from 2D experience, where pen/brush tilt only impacts curve aesthetic, and not shape.

\section{Sampling Target Curves for the User Study}
\label{app:sampling}

We wanted to design target curves that could be executed using a single smooth motion. Since users typically draw sharp corners using multiple strokes \cite{bae08ilovesketch}, we constrain our target curves to be smooth, created using cardinal cubic B-splines on the meshes, computed using Panozzo et al.~\shortcite{panozzo2013weighted}.
We also control the length and curvature complexity of the curves, as pilot testing showed that very simple and short curves can be reasonably executed by almost any projection technique. 
Curve length and complexity is modeled by placing spline control points at mesh vertices, and specifying the desired geodesic distance and Gau{\ss} map distance between consecutive control points on the curve. 

We represent a target curve using four parameters $\langle n, i_0, k_G, k_N \rangle$, where $n$ is the number of spline control points, $i_0$ the vertex index of the first control point, and $k_G, k_N$ constants that control the geodesic and normal map distance between consecutive control points. 
We define the desired geodesic distance between consecutive control points as, 
$D_G = k_G\times\|\mathtt{BBox}(\CM)\|$, where $\|\mathtt{BBox}(\CM)\|$ is the length of the  bounding box diagonal of $\CM$. The desired Gau{\ss} map distance (angle between the unit vertex normals) between consecutive control points is simply $k_N$.


A target curve $\bC_0, \ldots, \bC_{n-1}$  starting at vertex $\bv_{i_0}$ of the mesh is generated incrementally for $i>0$ as:
\begin{equation}
    \bC_i = \argmin_{\bv\in V'} \ \big(d_G(\bC_{i-1}, \bv) - D_G\big)^2 + \big(d_N(\bC_{i-1}, \bv) - k_N\big)^2,
\end{equation}
where $d_G$ and $d_N$ compute the geodesic and normal distance between two points on $\CM$, and $V'\subset V$ contains only those vertices of $\CM$ whose geodesic distance from $\bC_0, \ldots, \bC_{i-1}$ is at least $D_G/2$.
The restricted subset of vertices conveniently helps prevent (but doesn't fully avoid) self-intersecting or nearly self-intersecting curves.
Curves with complex self-intersections are less important practically, and can be particularly confusing for the curve re-creation task.
All our target curve samples were generated using $k_G\in[0.05, 0.25]$, $k_N\in[\pi/6, 5\pi/12]$, $n=6$, and a randomly chosen $i_0$.
The curves were manually inspected for self-intersections, and infringing curves rejected.

We then defined keypoints on the target curves as follows: curve endpoints were chosen as keypoints;
followed by greedily picking extrema of geodesic curvature, while ensuring that the arclength distance between any two consecutive keypoints was at least 3cm;
and concluding the procedure when the maximum arclength distance between any consecutive keypoints was below 15cm. Our target curves had between 4--9 keypoints (including endpoints).

\section{Additional Quantitative Analyses}
\label{app:quantitative}
\subsection{Quantifying Users' Tendency to Mimic}
\label{Sec:mimicriness}

The study also provided an opportunity to test if the users actually tended to mimic their intended curve $\CX$ in the mid-air stroke $\CP$. To quantify the ``mimcriness'' of a stroke, we subsample $\CP$ and $\CX$ into $m$ points as in \S~\ref{sec:strokeAccuracy}, use the correspondence as in Eq.~\ref{eq:distanceEquiparameter} and look at the variation in the distance (distance between the closest pair of corresponding points subtracted from that of the farthest pair) as a percentage of the target length $L_\CX$. We call this measure the \emph{mimicry violation} of a stroke. Intuitively, the lower the \emph{mimicry violation}, the closer the stroke $\CP$ is to being a perfect mimicry of $\CX$, going to zero if it is a precise translation of $\CX$. Notably, users depicted very similar trends to mimic for both the techniques---with 86\% (\emph{mimicry}), 80\% (\emph{spraycan}) strokes exhibiting \emph{mimicry violation} below 25\% of $L_\CX$, and 71\%, 66\% below 20\% of $L_\CX$---suggesting that mimicry is indeed a natural tendency.

\subsection{Consistency across Repeated Strokes}
\label{sec:repeat}

Recall that users repeated 2 of the 10 strokes per shape for both the techniques. To analyze consistency across the repeated strokes, we compared the values of the stroke accuracy measure $D_{eq}$ and the aesthetic measure $F_g$ between the original stroke and the corresponding repeated stroke. Specifically, we measured the relative change $|f(i) - f(i')|/f(i)$, where $(i, i')$ is a pair of original and repeated strokes, and $f(\cdot)$ is either $D_{eq}$ or $F_g$. Users were fairly consistent across both the techniques, with the average consistency for $D_{eq}$ being 35.4\% for \emph{mimicry} and 36.8\% for \emph{spraycan}, while for $F_g$, it was 36.5\% and 34.1\%, respectively. Note that the averages were computed after removing extreme outliers outside the $5\sigma$ threshold.

%% file: arxiv_mimicry.bbl

\begin{thebibliography}{80}


\ifx \showCODEN    \undefined \def \showCODEN     #1{\unskip}     \fi
\ifx \showDOI      \undefined \def \showDOI       #1{#1}\fi
\ifx \showISBNx    \undefined \def \showISBNx     #1{\unskip}     \fi
\ifx \showISBNxiii \undefined \def \showISBNxiii  #1{\unskip}     \fi
\ifx \showISSN     \undefined \def \showISSN      #1{\unskip}     \fi
\ifx \showLCCN     \undefined \def \showLCCN      #1{\unskip}     \fi
\ifx \shownote     \undefined \def \shownote      #1{#1}          \fi
\ifx \showarticletitle \undefined \def \showarticletitle #1{#1}   \fi
\ifx \showURL      \undefined \def \showURL       {\relax}        \fi
\providecommand\bibfield[2]{#2}
\providecommand\bibinfo[2]{#2}
\providecommand\natexlab[1]{#1}
\providecommand\showeprint[2][]{arXiv:#2}

\bibitem[\protect\citeauthoryear{Adobe}{Adobe}{2020}]%
        {adobe2020substance}
\bibfield{author}{\bibinfo{person}{Adobe}.} \bibinfo{year}{2020}\natexlab{}.
\newblock \bibinfo{title}{Substance Painter}.
\newblock
\newblock
\urldef\tempurl%
\url{https://www.substance3d.com/substance-painter/}
\showURL{%
\tempurl}


\bibitem[\protect\citeauthoryear{Adobe}{Adobe}{2021}]%
        {adobe2021medium}
\bibfield{author}{\bibinfo{person}{Adobe}.} \bibinfo{year}{2021}\natexlab{}.
\newblock \bibinfo{title}{Medium by Adobe}.
\newblock
\newblock
\urldef\tempurl%
\url{https://www.adobe.com/products/medium.html}
\showURL{%
\tempurl}


\bibitem[\protect\citeauthoryear{Andre and Saito}{Andre and Saito}{2011}]%
        {andre11single}
\bibfield{author}{\bibinfo{person}{Alexis Andre} {and} \bibinfo{person}{Suguru
  Saito}.} \bibinfo{year}{2011}\natexlab{}.
\newblock \showarticletitle{Single-View Sketch Based Modeling}. In
  \bibinfo{booktitle}{\emph{Proceedings of the Eighth Eurographics Symposium on
  Sketch-Based Interfaces and Modeling}} (Vancouver, British Columbia, Canada)
  \emph{(\bibinfo{series}{SBIM ’11})}. \bibinfo{publisher}{Association for
  Computing Machinery}, \bibinfo{address}{New York, NY, USA},
  \bibinfo{pages}{133--–140}.
\newblock
\showISBNx{9781450309066}
\urldef\tempurl%
\url{https://doi.org/10.1145/2021164.2021189}
\showDOI{\tempurl}


\bibitem[\protect\citeauthoryear{Arora, Kazi, Anderson, Grossman, Singh, and
  Fitzmaurice}{Arora et~al\mbox{.}}{2017}]%
        {arora2017experimental}
\bibfield{author}{\bibinfo{person}{Rahul Arora}, \bibinfo{person}{Rubaiat~Habib
  Kazi}, \bibinfo{person}{Fraser Anderson}, \bibinfo{person}{Tovi Grossman},
  \bibinfo{person}{Karan Singh}, {and} \bibinfo{person}{George Fitzmaurice}.}
  \bibinfo{year}{2017}\natexlab{}.
\newblock \showarticletitle{Experimental Evaluation of Sketching on Surfaces in
  VR}. In \bibinfo{booktitle}{\emph{Proceedings of the 2017 CHI Conference on
  Human Factors in Computing Systems}} (Denver, Colorado, USA)
  \emph{(\bibinfo{series}{CHI ’17})}. \bibinfo{publisher}{Association for
  Computing Machinery}, \bibinfo{address}{New York, NY, USA},
  \bibinfo{pages}{5643–5654}.
\newblock
\showISBNx{9781450346559}
\urldef\tempurl%
\url{https://doi.org/10.1145/3025453.3025474}
\showDOI{\tempurl}


\bibitem[\protect\citeauthoryear{Arora, Kazi, Grossman, Fitzmaurice, and
  Singh}{Arora et~al\mbox{.}}{2018}]%
        {arora2018symbiosis}
\bibfield{author}{\bibinfo{person}{Rahul Arora}, \bibinfo{person}{Rubaiat~Habib
  Kazi}, \bibinfo{person}{Tovi Grossman}, \bibinfo{person}{George Fitzmaurice},
  {and} \bibinfo{person}{Karan Singh}.} \bibinfo{year}{2018}\natexlab{}.
\newblock \showarticletitle{SymbiosisSketch: Combining 2D \& 3D Sketching for
  Designing Detailed 3D Objects in Situ}. In
  \bibinfo{booktitle}{\emph{Proceedings of the 2018 CHI Conference on Human
  Factors in Computing Systems}} (Montreal, Quebec, Canada)
  \emph{(\bibinfo{series}{CHI '18})}. \bibinfo{publisher}{ACM},
  \bibinfo{address}{New York, NY, USA}, \bibinfo{numpages}{15}~pages.
\newblock
\urldef\tempurl%
\url{https://doi.org/10.1145/3173574.3173759}
\showDOI{\tempurl}


\bibitem[\protect\citeauthoryear{Bae, Balakrishnan, and Singh}{Bae
  et~al\mbox{.}}{2008}]%
        {bae08ilovesketch}
\bibfield{author}{\bibinfo{person}{Seok-Hyung Bae}, \bibinfo{person}{Ravin
  Balakrishnan}, {and} \bibinfo{person}{Karan Singh}.}
  \bibinfo{year}{2008}\natexlab{}.
\newblock \showarticletitle{{ILoveSketch}: as-natural-as-possible sketching
  system for creating 3D curve models}. In
  \bibinfo{booktitle}{\emph{Proceedings of the 21st annual ACM symposium on
  User interface software and technology}} (Monterey, CA, USA)
  \emph{(\bibinfo{series}{UIST '08})}. \bibinfo{publisher}{ACM},
  \bibinfo{address}{New York, NY, USA}, \bibinfo{pages}{151--160}.
\newblock
\showISBNx{978-1-59593-975-3}
\urldef\tempurl%
\url{https://doi.org/10.1145/1449715.1449740}
\showDOI{\tempurl}


\bibitem[\protect\citeauthoryear{Barill, Dickson, Schmidt, Levin, and
  Jacobson}{Barill et~al\mbox{.}}{2018}]%
        {barill2018fast}
\bibfield{author}{\bibinfo{person}{Gavin Barill}, \bibinfo{person}{Neil~G.
  Dickson}, \bibinfo{person}{Ryan Schmidt}, \bibinfo{person}{David I.~W.
  Levin}, {and} \bibinfo{person}{Alec Jacobson}.}
  \bibinfo{year}{2018}\natexlab{}.
\newblock \showarticletitle{Fast Winding Numbers for Soups and Clouds}.
\newblock \bibinfo{journal}{\emph{ACM Trans. Graph.}} \bibinfo{volume}{37},
  \bibinfo{number}{4}, Article \bibinfo{articleno}{43} (\bibinfo{date}{July}
  \bibinfo{year}{2018}), \bibinfo{numpages}{12}~pages.
\newblock
\showISSN{0730-0301}
\urldef\tempurl%
\url{https://doi.org/10.1145/3197517.3201337}
\showDOI{\tempurl}


\bibitem[\protect\citeauthoryear{Cabral and Leedom}{Cabral and Leedom}{1993}]%
        {cabral1993imaging}
\bibfield{author}{\bibinfo{person}{Brian Cabral} {and}
  \bibinfo{person}{Leith~Casey Leedom}.} \bibinfo{year}{1993}\natexlab{}.
\newblock \showarticletitle{Imaging Vector Fields Using Line Integral
  Convolution}. In \bibinfo{booktitle}{\emph{Proceedings of the 20th Annual
  Conference on Computer Graphics and Interactive Techniques}} (Anaheim, CA)
  \emph{(\bibinfo{series}{SIGGRAPH ’93})}. \bibinfo{publisher}{Association
  for Computing Machinery}, \bibinfo{address}{New York, NY, USA},
  \bibinfo{pages}{263–270}.
\newblock
\showISBNx{0897916018}
\urldef\tempurl%
\url{https://doi.org/10.1145/166117.166151}
\showDOI{\tempurl}


\bibitem[\protect\citeauthoryear{Chen, Zhu, Shamir, Hu, and Cohen-Or}{Chen
  et~al\mbox{.}}{2013}]%
        {chen20133sweep}
\bibfield{author}{\bibinfo{person}{Tao Chen}, \bibinfo{person}{Zhe Zhu},
  \bibinfo{person}{Ariel Shamir}, \bibinfo{person}{Shi-Min Hu}, {and}
  \bibinfo{person}{Daniel Cohen-Or}.} \bibinfo{year}{2013}\natexlab{}.
\newblock \showarticletitle{3-Sweep: Extracting Editable Objects from a Single
  Photo}.
\newblock \bibinfo{journal}{\emph{ACM Trans. Graph.}} \bibinfo{volume}{32},
  \bibinfo{number}{6}, Article \bibinfo{articleno}{195} (\bibinfo{date}{Nov.}
  \bibinfo{year}{2013}), \bibinfo{numpages}{10}~pages.
\newblock
\showISSN{0730-0301}
\urldef\tempurl%
\url{https://doi.org/10.1145/2508363.2508378}
\showDOI{\tempurl}


\bibitem[\protect\citeauthoryear{Chen, Golovinskiy, and Funkhouser}{Chen
  et~al\mbox{.}}{2009}]%
        {chen2009benchmark}
\bibfield{author}{\bibinfo{person}{Xiaobai Chen}, \bibinfo{person}{Aleksey
  Golovinskiy}, {and} \bibinfo{person}{Thomas Funkhouser}.}
  \bibinfo{year}{2009}\natexlab{}.
\newblock \showarticletitle{A Benchmark for 3D Mesh Segmentation}.
\newblock \bibinfo{journal}{\emph{ACM Trans. Graph.}} \bibinfo{volume}{28},
  \bibinfo{number}{3}, Article \bibinfo{articleno}{73} (\bibinfo{date}{July}
  \bibinfo{year}{2009}), \bibinfo{numpages}{12}~pages.
\newblock
\showISSN{0730-0301}
\urldef\tempurl%
\url{https://doi.org/10.1145/1531326.1531379}
\showDOI{\tempurl}


\bibitem[\protect\citeauthoryear{Coleman and Singh}{Coleman and Singh}{2006}]%
        {coleman2006cords}
\bibfield{author}{\bibinfo{person}{Patrick Coleman} {and}
  \bibinfo{person}{Karan Singh}.} \bibinfo{year}{2006}\natexlab{}.
\newblock \showarticletitle{Cords: Geometric Curve Primitives for Modeling
  Contact}.
\newblock \bibinfo{journal}{\emph{IEEE Computer Graphics and Applications}}
  \bibinfo{volume}{26}, \bibinfo{number}{3} (\bibinfo{year}{2006}),
  \bibinfo{pages}{72--79}.
\newblock


\bibitem[\protect\citeauthoryear{Cox and Cox}{Cox and Cox}{2008}]%
        {cox2008multidimensional}
\bibfield{author}{\bibinfo{person}{Michael~AA Cox} {and}
  \bibinfo{person}{Trevor~F Cox}.} \bibinfo{year}{2008}\natexlab{}.
\newblock \showarticletitle{Multidimensional scaling}.
\newblock In \bibinfo{booktitle}{\emph{Handbook of data visualization}}.
  \bibinfo{publisher}{Springer}, \bibinfo{address}{New York, NY, USA},
  \bibinfo{pages}{315--347}.
\newblock


\bibitem[\protect\citeauthoryear{De~Paoli and Singh}{De~Paoli and
  Singh}{2015}]%
        {depauli2015secondskin}
\bibfield{author}{\bibinfo{person}{Chris De~Paoli} {and} \bibinfo{person}{Karan
  Singh}.} \bibinfo{year}{2015}\natexlab{}.
\newblock \showarticletitle{SecondSkin: Sketch-Based Construction of Layered 3D
  Models}.
\newblock \bibinfo{journal}{\emph{ACM Trans. Graph.}} \bibinfo{volume}{34},
  \bibinfo{number}{4}, Article \bibinfo{articleno}{126} (\bibinfo{date}{July}
  \bibinfo{year}{2015}), \bibinfo{numpages}{10}~pages.
\newblock
\showISSN{0730-0301}
\urldef\tempurl%
\url{https://doi.org/10.1145/2766948}
\showDOI{\tempurl}


\bibitem[\protect\citeauthoryear{Deering}{Deering}{1995}]%
        {deering1995holosketch}
\bibfield{author}{\bibinfo{person}{Michael~F Deering}.}
  \bibinfo{year}{1995}\natexlab{}.
\newblock \showarticletitle{HoloSketch: a virtual reality sketching/animation
  tool}.
\newblock \bibinfo{journal}{\emph{ACM Transactions on Computer-Human
  Interaction (TOCHI)}} \bibinfo{volume}{2}, \bibinfo{number}{3}
  (\bibinfo{year}{1995}), \bibinfo{pages}{220--238}.
\newblock


\bibitem[\protect\citeauthoryear{Fan, Wang, Xu, Deng, and Liu}{Fan
  et~al\mbox{.}}{2013}]%
        {fan2013modeling}
\bibfield{author}{\bibinfo{person}{Lubin Fan}, \bibinfo{person}{Ruimin Wang},
  \bibinfo{person}{Linlin Xu}, \bibinfo{person}{Jiansong Deng}, {and}
  \bibinfo{person}{Ligang Liu}.} \bibinfo{year}{2013}\natexlab{}.
\newblock \showarticletitle{Modeling by Drawing with Shadow Guidance}.
\newblock \bibinfo{journal}{\emph{Computer Graphics Forum}}
  \bibinfo{volume}{32}, \bibinfo{number}{7} (\bibinfo{year}{2013}),
  \bibinfo{pages}{157--166}.
\newblock
\urldef\tempurl%
\url{https://doi.org/10.1111/cgf.12223}
\showDOI{\tempurl}


\bibitem[\protect\citeauthoryear{Fan, Chi, Kaufman, and Oliveira}{Fan
  et~al\mbox{.}}{2004}]%
        {fan2004sketch}
\bibfield{author}{\bibinfo{person}{Zhe Fan}, \bibinfo{person}{Ma Chi},
  \bibinfo{person}{Arie Kaufman}, {and} \bibinfo{person}{Manuel~M. Oliveira}.}
  \bibinfo{year}{2004}\natexlab{}.
\newblock \showarticletitle{{A Sketch-Based Interface for Collaborative Design
  .}}. In \bibinfo{booktitle}{\emph{Sketch Based Interfaces and Modeling}}.
  \bibinfo{publisher}{The Eurographics Association}, \bibinfo{address}{Geneve,
  Switzerland}, \bibinfo{pages}{143--150}.
\newblock
\showISBNx{3-905673-16-9}
\showISSN{1812-3503}
\urldef\tempurl%
\url{https://doi.org/10.2312/SBM/SBM04/143-150}
\showDOI{\tempurl}


\bibitem[\protect\citeauthoryear{Fisher, Schr\"{o}der, Desbrun, and
  Hoppe}{Fisher et~al\mbox{.}}{2007}]%
        {fisher2007design}
\bibfield{author}{\bibinfo{person}{Matthew Fisher}, \bibinfo{person}{Peter
  Schr\"{o}der}, \bibinfo{person}{Mathieu Desbrun}, {and}
  \bibinfo{person}{Hugues Hoppe}.} \bibinfo{year}{2007}\natexlab{}.
\newblock \showarticletitle{Design of Tangent Vector Fields}.
\newblock \bibinfo{journal}{\emph{ACM Trans. Graph.}} \bibinfo{volume}{26},
  \bibinfo{number}{3} (\bibinfo{date}{July} \bibinfo{year}{2007}),
  \bibinfo{pages}{56–es}.
\newblock
\showISSN{0730-0301}
\urldef\tempurl%
\url{https://doi.org/10.1145/1276377.1276447}
\showDOI{\tempurl}


\bibitem[\protect\citeauthoryear{Fu, Wei, Tai, and Quan}{Fu
  et~al\mbox{.}}{2007}]%
        {fu2007sketching}
\bibfield{author}{\bibinfo{person}{Hongbo Fu}, \bibinfo{person}{Yichen Wei},
  \bibinfo{person}{Chiew-Lan Tai}, {and} \bibinfo{person}{Long Quan}.}
  \bibinfo{year}{2007}\natexlab{}.
\newblock \showarticletitle{Sketching Hairstyles}. In
  \bibinfo{booktitle}{\emph{Proceedings of the 4th Eurographics Workshop on
  Sketch-Based Interfaces and Modeling}} (Riverside, California)
  \emph{(\bibinfo{series}{SBIM ’07})}. \bibinfo{publisher}{Association for
  Computing Machinery}, \bibinfo{address}{New York, NY, USA},
  \bibinfo{pages}{31–36}.
\newblock
\showISBNx{9781595939135}
\urldef\tempurl%
\url{https://doi.org/10.1145/1384429.1384439}
\showDOI{\tempurl}


\bibitem[\protect\citeauthoryear{Gal, Sorkine, Mitra, and Cohen-Or}{Gal
  et~al\mbox{.}}{2009}]%
        {iwires}
\bibfield{author}{\bibinfo{person}{Ran Gal}, \bibinfo{person}{Olga Sorkine},
  \bibinfo{person}{Niloy~J. Mitra}, {and} \bibinfo{person}{Daniel Cohen-Or}.}
  \bibinfo{year}{2009}\natexlab{}.
\newblock \showarticletitle{iWIRES: An Analyze-and-Edit Approach to Shape
  Manipulation}.
\newblock \bibinfo{journal}{\emph{{ACM} Transactions on Graphics (Siggraph)}}
  \bibinfo{volume}{28}, \bibinfo{number}{3} (\bibinfo{year}{2009}),
  \bibinfo{pages}{\#33, 1--10}.
\newblock


\bibitem[\protect\citeauthoryear{Gehre, Bronstein, Kobbelt, and Solomon}{Gehre
  et~al\mbox{.}}{2018}]%
        {gehre2018interactive}
\bibfield{author}{\bibinfo{person}{Anne Gehre}, \bibinfo{person}{Michael
  Bronstein}, \bibinfo{person}{Leif Kobbelt}, {and} \bibinfo{person}{Justin
  Solomon}.} \bibinfo{year}{2018}\natexlab{}.
\newblock \showarticletitle{Interactive Curve Constrained Functional Maps}.
\newblock \bibinfo{journal}{\emph{Computer Graphics Forum}}
  \bibinfo{volume}{37}, \bibinfo{number}{5} (\bibinfo{year}{2018}),
  \bibinfo{pages}{1--12}.
\newblock
\urldef\tempurl%
\url{https://doi.org/10.1111/cgf.13486}
\showDOI{\tempurl}


\bibitem[\protect\citeauthoryear{Gooch and Gooch}{Gooch and Gooch}{2001}]%
        {goochbook}
\bibfield{author}{\bibinfo{person}{Bruce Gooch} {and} \bibinfo{person}{Amy
  Gooch}.} \bibinfo{year}{2001}\natexlab{}.
\newblock \bibinfo{booktitle}{\emph{Non-Photorealistic Rendering}}.
\newblock \bibinfo{publisher}{A. K. Peters}, \bibinfo{address}{USA}.
\newblock
\showISBNx{1568811330}


\bibitem[\protect\citeauthoryear{Google}{Google}{2020}]%
        {google2020tilt}
\bibfield{author}{\bibinfo{person}{Google}.} \bibinfo{year}{2020}\natexlab{}.
\newblock \bibinfo{title}{Tilt {Brush} by {Google}}.
\newblock
\newblock
\urldef\tempurl%
\url{https://www.tiltbrush.com/}
\showURL{%
\tempurl}


\bibitem[\protect\citeauthoryear{{Gravity Sketch}}{{Gravity Sketch}}{2020}]%
        {gravity2020gravity}
\bibfield{author}{\bibinfo{person}{{Gravity Sketch}}.}
  \bibinfo{year}{2020}\natexlab{}.
\newblock \bibinfo{title}{Gravity Sketch}.
\newblock
\newblock
\urldef\tempurl%
\url{https://www.gravitysketch.com/}
\showURL{%
\tempurl}


\bibitem[\protect\citeauthoryear{Grossman, Balakrishnan, Kurtenbach,
  Fitzmaurice, Khan, and Buxton}{Grossman et~al\mbox{.}}{2002}]%
        {grossman2002creating}
\bibfield{author}{\bibinfo{person}{Tovi Grossman}, \bibinfo{person}{Ravin
  Balakrishnan}, \bibinfo{person}{Gordon Kurtenbach}, \bibinfo{person}{George
  Fitzmaurice}, \bibinfo{person}{Azam Khan}, {and} \bibinfo{person}{Bill
  Buxton}.} \bibinfo{year}{2002}\natexlab{}.
\newblock \showarticletitle{Creating Principal 3D Curves with Digital Tape
  Drawing}. In \bibinfo{booktitle}{\emph{Proceedings of the SIGCHI Conference
  on Human Factors in Computing Systems}} (Minneapolis, Minnesota, USA)
  \emph{(\bibinfo{series}{CHI '02})}. \bibinfo{publisher}{ACM},
  \bibinfo{address}{New York, NY, USA}, \bibinfo{pages}{121--128}.
\newblock
\showISBNx{1-58113-453-3}
\urldef\tempurl%
\url{https://doi.org/10.1145/503376.503398}
\showDOI{\tempurl}


\bibitem[\protect\citeauthoryear{Heckel, Moltz, Tietjen, and Hahn}{Heckel
  et~al\mbox{.}}{2013}]%
        {heckel2013sketch}
\bibfield{author}{\bibinfo{person}{Frank Heckel}, \bibinfo{person}{Jan~H.
  Moltz}, \bibinfo{person}{Christian Tietjen}, {and} \bibinfo{person}{Horst~K.
  Hahn}.} \bibinfo{year}{2013}\natexlab{}.
\newblock \showarticletitle{Sketch-Based Editing Tools for Tumour Segmentation
  in 3D Medical Images}.
\newblock \bibinfo{journal}{\emph{Computer Graphics Forum}}
  \bibinfo{volume}{32}, \bibinfo{number}{8} (\bibinfo{year}{2013}),
  \bibinfo{pages}{144--157}.
\newblock
\urldef\tempurl%
\url{https://doi.org/10.1111/cgf.12193}
\showDOI{\tempurl}


\bibitem[\protect\citeauthoryear{Hertzmann and Zorin}{Hertzmann and
  Zorin}{2000}]%
        {hertzmann2003illustrating}
\bibfield{author}{\bibinfo{person}{Aaron Hertzmann} {and}
  \bibinfo{person}{Denis Zorin}.} \bibinfo{year}{2000}\natexlab{}.
\newblock \showarticletitle{Illustrating Smooth Surfaces}. In
  \bibinfo{booktitle}{\emph{Proceedings of the 27th Annual Conference on
  Computer Graphics and Interactive Techniques}}
  \emph{(\bibinfo{series}{SIGGRAPH ’00})}. \bibinfo{publisher}{ACM
  Press/Addison-Wesley Publishing Co.}, \bibinfo{address}{USA},
  \bibinfo{pages}{517–526}.
\newblock
\showISBNx{1581132085}
\urldef\tempurl%
\url{https://doi.org/10.1145/344779.345074}
\showDOI{\tempurl}


\bibitem[\protect\citeauthoryear{Hu, Zhou, Gao, Jacobson, Zorin, and
  Panozzo}{Hu et~al\mbox{.}}{2018}]%
        {hu2018tetwild}
\bibfield{author}{\bibinfo{person}{Yixin Hu}, \bibinfo{person}{Qingnan Zhou},
  \bibinfo{person}{Xifeng Gao}, \bibinfo{person}{Alec Jacobson},
  \bibinfo{person}{Denis Zorin}, {and} \bibinfo{person}{Daniele Panozzo}.}
  \bibinfo{year}{2018}\natexlab{}.
\newblock \showarticletitle{Tetrahedral Meshing in the Wild}.
\newblock \bibinfo{journal}{\emph{ACM Trans. Graph.}} \bibinfo{volume}{37},
  \bibinfo{number}{4}, Article \bibinfo{articleno}{60} (\bibinfo{date}{July}
  \bibinfo{year}{2018}), \bibinfo{numpages}{14}~pages.
\newblock
\showISSN{0730-0301}
\urldef\tempurl%
\url{https://doi.org/10.1145/3197517.3201353}
\showDOI{\tempurl}


\bibitem[\protect\citeauthoryear{Igarashi, Matsuoka, and Tanaka}{Igarashi
  et~al\mbox{.}}{1999}]%
        {igarashi1999teddy}
\bibfield{author}{\bibinfo{person}{Takeo Igarashi}, \bibinfo{person}{Satoshi
  Matsuoka}, {and} \bibinfo{person}{Hidehiko Tanaka}.}
  \bibinfo{year}{1999}\natexlab{}.
\newblock \showarticletitle{Teddy: A Sketching Interface for 3D Freeform
  Design}. In \bibinfo{booktitle}{\emph{Proceedings of the 26th Annual
  Conference on Computer Graphics and Interactive Techniques}}
  \emph{(\bibinfo{series}{SIGGRAPH ’99})}. \bibinfo{publisher}{ACM
  Press/Addison-Wesley Publishing Co.}, \bibinfo{address}{USA},
  \bibinfo{pages}{409--–416}.
\newblock
\showISBNx{0201485605}
\urldef\tempurl%
\url{https://doi.org/10.1145/311535.311602}
\showDOI{\tempurl}


\bibitem[\protect\citeauthoryear{Jackson and Keefe}{Jackson and Keefe}{2016}]%
        {jackson2016lift}
\bibfield{author}{\bibinfo{person}{Bret Jackson} {and}
  \bibinfo{person}{Daniel~F Keefe}.} \bibinfo{year}{2016}\natexlab{}.
\newblock \showarticletitle{Lift-off: Using Reference Imagery and Freehand
  Sketching to Create 3D Models in VR}.
\newblock \bibinfo{journal}{\emph{IEEE transactions on visualization and
  computer graphics}} \bibinfo{volume}{22}, \bibinfo{number}{4}
  (\bibinfo{year}{2016}), \bibinfo{pages}{1442--1451}.
\newblock


\bibitem[\protect\citeauthoryear{Jacobson, Panozzo, et~al\mbox{.}}{Jacobson
  et~al\mbox{.}}{2018}]%
        {libigl}
\bibfield{author}{\bibinfo{person}{Alec Jacobson}, \bibinfo{person}{Daniele
  Panozzo}, {et~al\mbox{.}}} \bibinfo{year}{2018}\natexlab{}.
\newblock \bibinfo{title}{{libigl}: A simple {C++} geometry processing
  library}.
\newblock
\newblock
\newblock
\shownote{https://libigl.github.io/.}


\bibitem[\protect\citeauthoryear{Jiang, Schneider, Zorin, and Panozzo}{Jiang
  et~al\mbox{.}}{2020}]%
        {jiang2020bijective}
\bibfield{author}{\bibinfo{person}{Zhongshi Jiang}, \bibinfo{person}{Teseo
  Schneider}, \bibinfo{person}{Denis Zorin}, {and} \bibinfo{person}{Daniele
  Panozzo}.} \bibinfo{year}{2020}\natexlab{}.
\newblock \showarticletitle{Bijective Projection in a Shell}.
\newblock \bibinfo{journal}{\emph{ACM Trans. Graph.}} \bibinfo{volume}{39},
  \bibinfo{number}{6}, Article \bibinfo{articleno}{247} (\bibinfo{date}{Nov.}
  \bibinfo{year}{2020}), \bibinfo{numpages}{18}~pages.
\newblock
\showISSN{0730-0301}
\urldef\tempurl%
\url{https://doi.org/10.1145/3414685.3417769}
\showDOI{\tempurl}


\bibitem[\protect\citeauthoryear{Jin, Song, Wang, Huang, Song, and He}{Jin
  et~al\mbox{.}}{2019}]%
        {jin2019shell}
\bibfield{author}{\bibinfo{person}{Yao Jin}, \bibinfo{person}{Dan Song},
  \bibinfo{person}{Tongtong Wang}, \bibinfo{person}{Jin Huang},
  \bibinfo{person}{Ying Song}, {and} \bibinfo{person}{Lili He}.}
  \bibinfo{year}{2019}\natexlab{}.
\newblock \showarticletitle{A shell space constrained approach for curve design
  on surface meshes}.
\newblock \bibinfo{journal}{\emph{Computer-Aided Design}}
  \bibinfo{volume}{113} (\bibinfo{year}{2019}), \bibinfo{pages}{24--34}.
\newblock
\showISSN{0010-4485}
\urldef\tempurl%
\url{https://doi.org/10.1016/j.cad.2019.03.001}
\showDOI{\tempurl}


\bibitem[\protect\citeauthoryear{Jung, Gross, and Do}{Jung
  et~al\mbox{.}}{2002}]%
        {jung2002annotating}
\bibfield{author}{\bibinfo{person}{Thomas Jung}, \bibinfo{person}{Mark~D.
  Gross}, {and} \bibinfo{person}{Ellen Yi-Luen Do}.}
  \bibinfo{year}{2002}\natexlab{}.
\newblock \showarticletitle{Annotating and Sketching on 3D Web Models}. In
  \bibinfo{booktitle}{\emph{Proceedings of the 7th International Conference on
  Intelligent User Interfaces}} (San Francisco, California, USA)
  \emph{(\bibinfo{series}{IUI ’02})}. \bibinfo{publisher}{Association for
  Computing Machinery}, \bibinfo{address}{New York, NY, USA},
  \bibinfo{pages}{95–102}.
\newblock
\showISBNx{1581134592}
\urldef\tempurl%
\url{https://doi.org/10.1145/502716.502733}
\showDOI{\tempurl}


\bibitem[\protect\citeauthoryear{Kalnins, Markosian, Meier, Kowalski, Lee,
  Davidson, Webb, Hughes, and Finkelstein}{Kalnins et~al\mbox{.}}{2002}]%
        {kalnins2002wysiwyg}
\bibfield{author}{\bibinfo{person}{Robert~D. Kalnins}, \bibinfo{person}{Lee
  Markosian}, \bibinfo{person}{Barbara~J. Meier}, \bibinfo{person}{Michael~A.
  Kowalski}, \bibinfo{person}{Joseph~C. Lee}, \bibinfo{person}{Philip~L.
  Davidson}, \bibinfo{person}{Matthew Webb}, \bibinfo{person}{John~F. Hughes},
  {and} \bibinfo{person}{Adam Finkelstein}.} \bibinfo{year}{2002}\natexlab{}.
\newblock \showarticletitle{WYSIWYG NPR: Drawing Strokes Directly on 3D
  Models}.
\newblock \bibinfo{journal}{\emph{ACM Trans. Graph.}} \bibinfo{volume}{21},
  \bibinfo{number}{3} (\bibinfo{date}{July} \bibinfo{year}{2002}),
  \bibinfo{pages}{755–--762}.
\newblock
\showISSN{0730-0301}
\urldef\tempurl%
\url{https://doi.org/10.1145/566654.566648}
\showDOI{\tempurl}


\bibitem[\protect\citeauthoryear{Kamuro, Minamizawa, and Tachi}{Kamuro
  et~al\mbox{.}}{2011}]%
        {kamuro20113d}
\bibfield{author}{\bibinfo{person}{Sho Kamuro}, \bibinfo{person}{Kouta
  Minamizawa}, {and} \bibinfo{person}{Susumu Tachi}.}
  \bibinfo{year}{2011}\natexlab{}.
\newblock \showarticletitle{3D Haptic Modeling System using Ungrounded
  Pen-shaped Kinesthetic Display}. In \bibinfo{booktitle}{\emph{2011 IEEE
  Virtual Reality Conference}}. \bibinfo{publisher}{IEEE},
  \bibinfo{address}{New York, NY, USA}, \bibinfo{pages}{217--218}.
\newblock


\bibitem[\protect\citeauthoryear{Kara and Shimada}{Kara and Shimada}{2007}]%
        {kara2007sketch}
\bibfield{author}{\bibinfo{person}{Levent~Burak Kara} {and}
  \bibinfo{person}{Kenji Shimada}.} \bibinfo{year}{2007}\natexlab{}.
\newblock \showarticletitle{Sketch-Based 3D-Shape Creation for Industrial
  Styling Design}.
\newblock \bibinfo{journal}{\emph{IEEE Comput. Graph. Appl.}}
  \bibinfo{volume}{27}, \bibinfo{number}{1} (\bibinfo{date}{Jan.}
  \bibinfo{year}{2007}), \bibinfo{pages}{60--71}.
\newblock
\showISSN{0272-1716}
\urldef\tempurl%
\url{https://doi.org/10.1109/MCG.2007.18}
\showDOI{\tempurl}


\bibitem[\protect\citeauthoryear{Keefe, Zeleznik, and Laidlaw}{Keefe
  et~al\mbox{.}}{2007}]%
        {keefe2007drawing}
\bibfield{author}{\bibinfo{person}{Daniel Keefe}, \bibinfo{person}{Robert
  Zeleznik}, {and} \bibinfo{person}{David Laidlaw}.}
  \bibinfo{year}{2007}\natexlab{}.
\newblock \showarticletitle{Drawing on Air: Input Techniques for Controlled 3D
  Line Illustration}.
\newblock \bibinfo{journal}{\emph{IEEE Transactions on Visualization and
  Computer Graphics}} \bibinfo{volume}{13}, \bibinfo{number}{5}
  (\bibinfo{year}{2007}), \bibinfo{pages}{1067--1081}.
\newblock


\bibitem[\protect\citeauthoryear{Keefe, Feliz, Moscovich, Laidlaw, and
  LaViola}{Keefe et~al\mbox{.}}{2001}]%
        {keefe2001cavepainting}
\bibfield{author}{\bibinfo{person}{Daniel~F. Keefe},
  \bibinfo{person}{Daniel~Acevedo Feliz}, \bibinfo{person}{Tomer Moscovich},
  \bibinfo{person}{David~H. Laidlaw}, {and} \bibinfo{person}{Joseph~J.
  LaViola}.} \bibinfo{year}{2001}\natexlab{}.
\newblock \showarticletitle{CavePainting: A Fully Immersive 3D Artistic Medium
  and Interactive Experience}. In \bibinfo{booktitle}{\emph{Proceedings of the
  2001 Symposium on Interactive 3D Graphics}} \emph{(\bibinfo{series}{I3D
  ’01})}. \bibinfo{publisher}{Association for Computing Machinery},
  \bibinfo{address}{New York, NY, USA}, \bibinfo{pages}{85–93}.
\newblock
\showISBNx{1581132921}
\urldef\tempurl%
\url{https://doi.org/10.1145/364338.364370}
\showDOI{\tempurl}


\bibitem[\protect\citeauthoryear{Krishnamurthy and Levoy}{Krishnamurthy and
  Levoy}{1996}]%
        {krishlevoy}
\bibfield{author}{\bibinfo{person}{Venkat Krishnamurthy} {and}
  \bibinfo{person}{Marc Levoy}.} \bibinfo{year}{1996}\natexlab{}.
\newblock \showarticletitle{Fitting Smooth Surfaces to Dense Polygon Meshes}.
  In \bibinfo{booktitle}{\emph{Proceedings of the 23rd Annual Conference on
  Computer Graphics and Interactive Techniques}}
  \emph{(\bibinfo{series}{SIGGRAPH '96})}. \bibinfo{publisher}{Association for
  Computing Machinery}, \bibinfo{address}{New York, NY, USA},
  \bibinfo{pages}{313–324}.
\newblock
\showISBNx{0897917464}
\urldef\tempurl%
\url{https://doi.org/10.1145/237170.237270}
\showDOI{\tempurl}


\bibitem[\protect\citeauthoryear{Krs, Yumer, Carr, Benes, and M\v{e}ch}{Krs
  et~al\mbox{.}}{2017}]%
        {krs2017Skippy}
\bibfield{author}{\bibinfo{person}{Vojt\v{e}ch Krs}, \bibinfo{person}{Ersin
  Yumer}, \bibinfo{person}{Nathan Carr}, \bibinfo{person}{Bedrich Benes}, {and}
  \bibinfo{person}{Radom\'{\i}r M\v{e}ch}.} \bibinfo{year}{2017}\natexlab{}.
\newblock \showarticletitle{Skippy: Single View 3D Curve Interactive Modeling}.
\newblock \bibinfo{journal}{\emph{ACM Trans. Graph.}} \bibinfo{volume}{36},
  \bibinfo{number}{4}, Article \bibinfo{articleno}{128} (\bibinfo{date}{July}
  \bibinfo{year}{2017}), \bibinfo{numpages}{12}~pages.
\newblock
\showISSN{0730-0301}
\urldef\tempurl%
\url{https://doi.org/10.1145/3072959.3073603}
\showDOI{\tempurl}


\bibitem[\protect\citeauthoryear{Kwan and Fu}{Kwan and Fu}{2019}]%
        {kwan2019mobi}
\bibfield{author}{\bibinfo{person}{Kin~Chung Kwan} {and}
  \bibinfo{person}{Hongbo Fu}.} \bibinfo{year}{2019}\natexlab{}.
\newblock \showarticletitle{Mobi3DSketch: 3D Sketching in Mobile AR}. In
  \bibinfo{booktitle}{\emph{Proceedings of the 2019 CHI Conference on Human
  Factors in Computing Systems}} (Glasgow, Scotland, UK)
  \emph{(\bibinfo{series}{CHI '19})}. \bibinfo{publisher}{Association for
  Computing Machinery}, \bibinfo{address}{New York, NY, USA},
  \bibinfo{pages}{1–11}.
\newblock
\showISBNx{9781450359702}
\urldef\tempurl%
\url{https://doi.org/10.1145/3290605.3300406}
\showDOI{\tempurl}


\bibitem[\protect\citeauthoryear{Lefebvre and Hoppe}{Lefebvre and
  Hoppe}{2006}]%
        {lefebre2006appearance}
\bibfield{author}{\bibinfo{person}{Sylvain Lefebvre} {and}
  \bibinfo{person}{Hugues Hoppe}.} \bibinfo{year}{2006}\natexlab{}.
\newblock \showarticletitle{Appearance-Space Texture Synthesis}.
\newblock \bibinfo{journal}{\emph{ACM Trans. Graph.}} \bibinfo{volume}{25},
  \bibinfo{number}{3} (\bibinfo{date}{July} \bibinfo{year}{2006}),
  \bibinfo{pages}{541–548}.
\newblock
\showISSN{0730-0301}
\urldef\tempurl%
\url{https://doi.org/10.1145/1141911.1141921}
\showDOI{\tempurl}


\bibitem[\protect\citeauthoryear{L\'{e}vy, Petitjean, Ray, and
  Maillot}{L\'{e}vy et~al\mbox{.}}{2002}]%
        {levy2002least}
\bibfield{author}{\bibinfo{person}{Bruno L\'{e}vy}, \bibinfo{person}{Sylvain
  Petitjean}, \bibinfo{person}{Nicolas Ray}, {and} \bibinfo{person}{J\'{e}rome
  Maillot}.} \bibinfo{year}{2002}\natexlab{}.
\newblock \showarticletitle{Least Squares Conformal Maps for Automatic Texture
  Atlas Generation}.
\newblock \bibinfo{journal}{\emph{ACM Trans. Graph.}} \bibinfo{volume}{21},
  \bibinfo{number}{3} (\bibinfo{date}{July} \bibinfo{year}{2002}),
  \bibinfo{pages}{362–371}.
\newblock
\showISSN{0730-0301}
\urldef\tempurl%
\url{https://doi.org/10.1145/566654.566590}
\showDOI{\tempurl}


\bibitem[\protect\citeauthoryear{Machuca, Asente, Stuerzlinger, Lu, and
  Kim}{Machuca et~al\mbox{.}}{2018}]%
        {machuca2018multi}
\bibfield{author}{\bibinfo{person}{Mayra D.~Barrera Machuca},
  \bibinfo{person}{Paul Asente}, \bibinfo{person}{Wolfgang Stuerzlinger},
  \bibinfo{person}{Jingwan Lu}, {and} \bibinfo{person}{Byungmoon Kim}.}
  \bibinfo{year}{2018}\natexlab{}.
\newblock \showarticletitle{Multiplanes: Assisted Freehand VR Sketching}. In
  \bibinfo{booktitle}{\emph{Proceedings of the Symposium on Spatial User
  Interaction}} (Berlin, Germany) \emph{(\bibinfo{series}{SUI '18})}.
  \bibinfo{publisher}{Association for Computing Machinery},
  \bibinfo{address}{New York, NY, USA}, \bibinfo{pages}{36–47}.
\newblock
\showISBNx{9781450357081}
\urldef\tempurl%
\url{https://doi.org/10.1145/3267782.3267786}
\showDOI{\tempurl}


\bibitem[\protect\citeauthoryear{Machuca, Stuerzlinger, and Asente}{Machuca
  et~al\mbox{.}}{2019}]%
        {machuca2019effect}
\bibfield{author}{\bibinfo{person}{Mayra Donaji~Barrera Machuca},
  \bibinfo{person}{Wolfgang Stuerzlinger}, {and} \bibinfo{person}{Paul
  Asente}.} \bibinfo{year}{2019}\natexlab{}.
\newblock \showarticletitle{The Effect of Spatial Ability on Immersive 3D
  Drawing}. In \bibinfo{booktitle}{\emph{Proceedings of the ACM Conference on
  Creativity \& Cognition (C\&C’19)}}. \bibinfo{publisher}{ACM},
  \bibinfo{address}{New York, NY, USA}, \bibinfo{pages}{173–186}.
\newblock


\bibitem[\protect\citeauthoryear{McCrae and Singh}{McCrae and Singh}{2008}]%
        {mccrae2008sketching}
\bibfield{author}{\bibinfo{person}{James McCrae} {and} \bibinfo{person}{Karan
  Singh}.} \bibinfo{year}{2008}\natexlab{}.
\newblock \showarticletitle{Sketching Piecewise Clothoid Curves}. In
  \bibinfo{booktitle}{\emph{Proceedings of the Fifth Eurographics Conference on
  Sketch-Based Interfaces and Modeling}} (Annecy, France)
  \emph{(\bibinfo{series}{SBM’08})}. \bibinfo{publisher}{Eurographics
  Association}, \bibinfo{address}{Goslar, DEU}, \bibinfo{pages}{1–8}.
\newblock
\showISBNx{9783905674071}


\bibitem[\protect\citeauthoryear{McCrae, Singh, and Mitra}{McCrae
  et~al\mbox{.}}{2011}]%
        {mccrae2011slices}
\bibfield{author}{\bibinfo{person}{James McCrae}, \bibinfo{person}{Karan
  Singh}, {and} \bibinfo{person}{Niloy~J. Mitra}.}
  \bibinfo{year}{2011}\natexlab{}.
\newblock \showarticletitle{Slices: A Shape-Proxy Based on Planar Sections}.
\newblock \bibinfo{journal}{\emph{ACM Trans. Graph.}} \bibinfo{volume}{30},
  \bibinfo{number}{6} (\bibinfo{date}{Dec.} \bibinfo{year}{2011}),
  \bibinfo{pages}{1–12}.
\newblock
\showISSN{0730-0301}
\urldef\tempurl%
\url{https://doi.org/10.1145/2070781.2024202}
\showDOI{\tempurl}


\bibitem[\protect\citeauthoryear{McCrae, Umetani, and Singh}{McCrae
  et~al\mbox{.}}{2014}]%
        {flatfab}
\bibfield{author}{\bibinfo{person}{James McCrae}, \bibinfo{person}{Nobuyuki
  Umetani}, {and} \bibinfo{person}{Karan Singh}.}
  \bibinfo{year}{2014}\natexlab{}.
\newblock \showarticletitle{FlatFitFab: Interactive Modeling with Planar
  Sections}. In \bibinfo{booktitle}{\emph{Proceedings of the 27th Annual ACM
  Symposium on User Interface Software and Technology}} (Honolulu, Hawaii, USA)
  \emph{(\bibinfo{series}{UIST '14})}. \bibinfo{publisher}{Association for
  Computing Machinery}, \bibinfo{address}{New York, NY, USA},
  \bibinfo{pages}{13–22}.
\newblock
\showISBNx{9781450330695}
\urldef\tempurl%
\url{https://doi.org/10.1145/2642918.2647388}
\showDOI{\tempurl}


\bibitem[\protect\citeauthoryear{Meng, Fan, and Liu}{Meng
  et~al\mbox{.}}{2011}]%
        {meng2011icutter}
\bibfield{author}{\bibinfo{person}{Min Meng}, \bibinfo{person}{Lubin Fan},
  {and} \bibinfo{person}{Ligang Liu}.} \bibinfo{year}{2011}\natexlab{}.
\newblock \showarticletitle{iCutter: A Direct Cut-out Tool for 3D Shapes}.
\newblock \bibinfo{journal}{\emph{Computer Animation and Virtual Worlds}}
  \bibinfo{volume}{22}, \bibinfo{number}{4} (\bibinfo{year}{2011}),
  \bibinfo{pages}{335--342}.
\newblock
\urldef\tempurl%
\url{https://doi.org/10.1002/cav.422}
\showDOI{\tempurl}


\bibitem[\protect\citeauthoryear{Nealen, Igarashi, Sorkine, and Alexa}{Nealen
  et~al\mbox{.}}{2007}]%
        {nealen2007fibermesh}
\bibfield{author}{\bibinfo{person}{Andrew Nealen}, \bibinfo{person}{Takeo
  Igarashi}, \bibinfo{person}{Olga Sorkine}, {and} \bibinfo{person}{Marc
  Alexa}.} \bibinfo{year}{2007}\natexlab{}.
\newblock \showarticletitle{FiberMesh: Designing Freeform Surfaces with 3D
  Curves}.
\newblock \bibinfo{journal}{\emph{ACM Trans. Graph.}} \bibinfo{volume}{26},
  \bibinfo{number}{3} (\bibinfo{date}{July} \bibinfo{year}{2007}),
  \bibinfo{pages}{41–es}.
\newblock
\showISSN{0730-0301}
\urldef\tempurl%
\url{https://doi.org/10.1145/1276377.1276429}
\showDOI{\tempurl}


\bibitem[\protect\citeauthoryear{Oculus}{Oculus}{2020}]%
        {oculus2020quill}
\bibfield{author}{\bibinfo{person}{Oculus}.} \bibinfo{year}{2020}\natexlab{}.
\newblock \bibinfo{title}{Quill}.
\newblock
\newblock
\urldef\tempurl%
\url{https://www.oculus.com/experiences/rift/1118609381580656/}
\showURL{%
\tempurl}


\bibitem[\protect\citeauthoryear{Olsen, Samavati, Sousa, and Jorge}{Olsen
  et~al\mbox{.}}{2009}]%
        {olsen2009sketch}
\bibfield{author}{\bibinfo{person}{Luke Olsen}, \bibinfo{person}{Faramarz~F.
  Samavati}, \bibinfo{person}{Mario~Costa Sousa}, {and}
  \bibinfo{person}{Joaquim~A. Jorge}.} \bibinfo{year}{2009}\natexlab{}.
\newblock \showarticletitle{Sketch-based Modeling: A survey}.
\newblock \bibinfo{journal}{\emph{Computers and Graphics}}
  \bibinfo{volume}{33}, \bibinfo{number}{1} (\bibinfo{year}{2009}),
  \bibinfo{pages}{85--103}.
\newblock
\showISSN{0097-8493}
\urldef\tempurl%
\url{https://doi.org/10.1016/j.cag.2008.09.013}
\showDOI{\tempurl}


\bibitem[\protect\citeauthoryear{Ortega and Vincent}{Ortega and
  Vincent}{2014}]%
        {ortega2014direct}
\bibfield{author}{\bibinfo{person}{Micha\"{e}l Ortega} {and}
  \bibinfo{person}{Thomas Vincent}.} \bibinfo{year}{2014}\natexlab{}.
\newblock \showarticletitle{Direct Drawing on 3D Shapes with Automated Camera
  Control}. In \bibinfo{booktitle}{\emph{Proceedings of the SIGCHI Conference
  on Human Factors in Computing Systems}} (Toronto, Canada)
  \emph{(\bibinfo{series}{CHI ’14})}. \bibinfo{publisher}{Association for
  Computing Machinery}, \bibinfo{address}{New York, NY, USA},
  \bibinfo{pages}{2047–2050}.
\newblock
\showISBNx{9781450324731}
\urldef\tempurl%
\url{https://doi.org/10.1145/2556288.2557242}
\showDOI{\tempurl}


\bibitem[\protect\citeauthoryear{Paczkowski, Kim, Morvan, Dorsey, Rushmeier,
  and O'Sullivan}{Paczkowski et~al\mbox{.}}{2011}]%
        {paczkowski2011insitu}
\bibfield{author}{\bibinfo{person}{Patrick Paczkowski}, \bibinfo{person}{Min~H.
  Kim}, \bibinfo{person}{Yann Morvan}, \bibinfo{person}{Julie Dorsey},
  \bibinfo{person}{Holly Rushmeier}, {and} \bibinfo{person}{Carol O'Sullivan}.}
  \bibinfo{year}{2011}\natexlab{}.
\newblock \showarticletitle{Insitu: Sketching Architectural Designs in
  Context}.
\newblock \bibinfo{journal}{\emph{ACM Trans. Graph.}} \bibinfo{volume}{30},
  \bibinfo{number}{6} (\bibinfo{date}{Dec.} \bibinfo{year}{2011}),
  \bibinfo{pages}{1–10}.
\newblock
\showISSN{0730-0301}
\urldef\tempurl%
\url{https://doi.org/10.1145/2070781.2024216}
\showDOI{\tempurl}


\bibitem[\protect\citeauthoryear{Panozzo, Baran, Diamanti, and
  Sorkine-Hornung}{Panozzo et~al\mbox{.}}{2013}]%
        {panozzo2013weighted}
\bibfield{author}{\bibinfo{person}{Daniele Panozzo}, \bibinfo{person}{Ilya
  Baran}, \bibinfo{person}{Olga Diamanti}, {and} \bibinfo{person}{Olga
  Sorkine-Hornung}.} \bibinfo{year}{2013}\natexlab{}.
\newblock \showarticletitle{Weighted Averages on Surfaces}.
\newblock \bibinfo{journal}{\emph{ACM Trans. Graph.}} \bibinfo{volume}{32},
  \bibinfo{number}{4}, Article \bibinfo{articleno}{60} (\bibinfo{date}{July}
  \bibinfo{year}{2013}), \bibinfo{numpages}{12}~pages.
\newblock
\showISSN{0730-0301}
\urldef\tempurl%
\url{https://doi.org/10.1145/2461912.2461935}
\showDOI{\tempurl}


\bibitem[\protect\citeauthoryear{Polthier and Schmies}{Polthier and
  Schmies}{2006}]%
        {polthier2006straightest}
\bibfield{author}{\bibinfo{person}{Konrad Polthier} {and}
  \bibinfo{person}{Markus Schmies}.} \bibinfo{year}{2006}\natexlab{}.
\newblock \showarticletitle{Straightest Geodesics on Polyhedral Surfaces}. In
  \bibinfo{booktitle}{\emph{ACM SIGGRAPH 2006 Courses}} (Boston, Massachusetts)
  \emph{(\bibinfo{series}{SIGGRAPH ’06})}. \bibinfo{publisher}{Association
  for Computing Machinery}, \bibinfo{address}{New York, NY, USA},
  \bibinfo{pages}{30–38}.
\newblock
\showISBNx{1595933646}
\urldef\tempurl%
\url{https://doi.org/10.1145/1185657.1185664}
\showDOI{\tempurl}


\bibitem[\protect\citeauthoryear{Rabinovich, Poranne, Panozzo, and
  Sorkine-Hornung}{Rabinovich et~al\mbox{.}}{2017}]%
        {rabinovich2017scalable}
\bibfield{author}{\bibinfo{person}{Michael Rabinovich}, \bibinfo{person}{Roi
  Poranne}, \bibinfo{person}{Daniele Panozzo}, {and} \bibinfo{person}{Olga
  Sorkine-Hornung}.} \bibinfo{year}{2017}\natexlab{}.
\newblock \showarticletitle{Scalable Locally Injective Mappings}.
\newblock \bibinfo{journal}{\emph{ACM Trans. Graph.}} \bibinfo{volume}{36},
  \bibinfo{number}{2}, Article \bibinfo{articleno}{16} (\bibinfo{date}{April}
  \bibinfo{year}{2017}), \bibinfo{numpages}{16}~pages.
\newblock
\showISSN{0730-0301}
\urldef\tempurl%
\url{https://doi.org/10.1145/2983621}
\showDOI{\tempurl}


\bibitem[\protect\citeauthoryear{Sawhney and Crane}{Sawhney and Crane}{2017}]%
        {sawhney2017boundary}
\bibfield{author}{\bibinfo{person}{Rohan Sawhney} {and} \bibinfo{person}{Keenan
  Crane}.} \bibinfo{year}{2017}\natexlab{}.
\newblock \showarticletitle{Boundary First Flattening}.
\newblock \bibinfo{journal}{\emph{ACM Trans. Graph.}} \bibinfo{volume}{37},
  \bibinfo{number}{1}, Article \bibinfo{articleno}{5} (\bibinfo{date}{Dec.}
  \bibinfo{year}{2017}), \bibinfo{numpages}{14}~pages.
\newblock
\showISSN{0730-0301}
\urldef\tempurl%
\url{https://doi.org/10.1145/3132705}
\showDOI{\tempurl}


\bibitem[\protect\citeauthoryear{Schkolne, Pruett, and Schr{\"o}der}{Schkolne
  et~al\mbox{.}}{2001}]%
        {schkolne2001surface}
\bibfield{author}{\bibinfo{person}{Steven Schkolne}, \bibinfo{person}{Michael
  Pruett}, {and} \bibinfo{person}{Peter Schr{\"o}der}.}
  \bibinfo{year}{2001}\natexlab{}.
\newblock \showarticletitle{Surface Drawing: Creating Organic 3D Shapes with
  the Hand and Tangible Tools}. In \bibinfo{booktitle}{\emph{Proceedings of the
  SIGCHI conference on Human factors in computing systems}}.
  \bibinfo{publisher}{ACM}, \bibinfo{address}{New York, NY, USA},
  \bibinfo{pages}{261--268}.
\newblock


\bibitem[\protect\citeauthoryear{Schmid, Senn, Gross, and Sumner}{Schmid
  et~al\mbox{.}}{2011}]%
        {schmidt2011overcoat}
\bibfield{author}{\bibinfo{person}{Johannes Schmid},
  \bibinfo{person}{Martin~Sebastian Senn}, \bibinfo{person}{Markus Gross},
  {and} \bibinfo{person}{Robert~W. Sumner}.} \bibinfo{year}{2011}\natexlab{}.
\newblock \showarticletitle{OverCoat: An Implicit Canvas for 3D Painting}.
\newblock \bibinfo{journal}{\emph{ACM Trans. Graph.}} \bibinfo{volume}{30},
  \bibinfo{number}{4}, Article \bibinfo{articleno}{28} (\bibinfo{date}{July}
  \bibinfo{year}{2011}), \bibinfo{numpages}{10}~pages.
\newblock
\showISSN{0730-0301}
\urldef\tempurl%
\url{https://doi.org/10.1145/2010324.1964923}
\showDOI{\tempurl}


\bibitem[\protect\citeauthoryear{Schmidt}{Schmidt}{2017}]%
        {geometry3sharp}
\bibfield{author}{\bibinfo{person}{Ryan Schmidt}.}
  \bibinfo{year}{2017}\natexlab{}.
\newblock \bibinfo{title}{{geometry3sharp}: Open-Source (Boost-license) C\#
  Library for Geometric Computing}.
\newblock
\newblock
\newblock
\shownote{https://github.com/gradientspace/geometry3Sharp.}


\bibitem[\protect\citeauthoryear{Schmidt, Grimm, and Wyvill}{Schmidt
  et~al\mbox{.}}{2006}]%
        {schmidt2006discrete}
\bibfield{author}{\bibinfo{person}{Ryan Schmidt}, \bibinfo{person}{Cindy
  Grimm}, {and} \bibinfo{person}{Brian Wyvill}.}
  \bibinfo{year}{2006}\natexlab{}.
\newblock \showarticletitle{Interactive Decal Compositing with Discrete
  Exponential Maps}.
\newblock \bibinfo{journal}{\emph{ACM Trans. Graph.}} \bibinfo{volume}{25},
  \bibinfo{number}{3} (\bibinfo{date}{July} \bibinfo{year}{2006}),
  \bibinfo{pages}{605–613}.
\newblock
\showISSN{0730-0301}
\urldef\tempurl%
\url{https://doi.org/10.1145/1141911.1141930}
\showDOI{\tempurl}


\bibitem[\protect\citeauthoryear{Schmidt, Khan, Singh, and Kurtenbach}{Schmidt
  et~al\mbox{.}}{2009}]%
        {schmidt2009analytic}
\bibfield{author}{\bibinfo{person}{Ryan Schmidt}, \bibinfo{person}{Azam Khan},
  \bibinfo{person}{Karan Singh}, {and} \bibinfo{person}{Gord Kurtenbach}.}
  \bibinfo{year}{2009}\natexlab{}.
\newblock \showarticletitle{Analytic Drawing of 3D Scaffolds}. In
  \bibinfo{booktitle}{\emph{ACM SIGGRAPH Asia 2009 Papers}} (Yokohama, Japan)
  \emph{(\bibinfo{series}{SIGGRAPH Asia ’09})}.
  \bibinfo{publisher}{Association for Computing Machinery},
  \bibinfo{address}{New York, NY, USA}, Article \bibinfo{articleno}{149},
  \bibinfo{numpages}{10}~pages.
\newblock
\showISBNx{9781605588582}
\urldef\tempurl%
\url{https://doi.org/10.1145/1661412.1618495}
\showDOI{\tempurl}


\bibitem[\protect\citeauthoryear{Schmidt and Singh}{Schmidt and Singh}{2010}]%
        {schmidt2010meshmixer}
\bibfield{author}{\bibinfo{person}{Ryan Schmidt} {and} \bibinfo{person}{Karan
  Singh}.} \bibinfo{year}{2010}\natexlab{}.
\newblock \showarticletitle{Meshmixer: An Interface for Rapid Mesh
  Composition}. In \bibinfo{booktitle}{\emph{ACM SIGGRAPH 2010 Talks}} (Los
  Angeles, California) \emph{(\bibinfo{series}{SIGGRAPH '10})}.
  \bibinfo{publisher}{ACM}, \bibinfo{address}{New York, NY, USA}, Article
  \bibinfo{articleno}{6}, \bibinfo{numpages}{1}~pages.
\newblock
\showISBNx{978-1-4503-0394-1}
\urldef\tempurl%
\url{https://doi.org/10.1145/1837026.1837034}
\showDOI{\tempurl}


\bibitem[\protect\citeauthoryear{Shao, Bousseau, Sheffer, and Singh}{Shao
  et~al\mbox{.}}{2012}]%
        {shao2012crossshade}
\bibfield{author}{\bibinfo{person}{Cloud Shao}, \bibinfo{person}{Adrien
  Bousseau}, \bibinfo{person}{Alla Sheffer}, {and} \bibinfo{person}{Karan
  Singh}.} \bibinfo{year}{2012}\natexlab{}.
\newblock \showarticletitle{CrossShade: Shading Concept Sketches Using
  Cross-section Curves}.
\newblock \bibinfo{journal}{\emph{ACM Trans. Graph.}} \bibinfo{volume}{31},
  \bibinfo{number}{4}, Article \bibinfo{articleno}{45} (\bibinfo{date}{July}
  \bibinfo{year}{2012}), \bibinfo{numpages}{11}~pages.
\newblock
\showISSN{0730-0301}
\urldef\tempurl%
\url{https://doi.org/10.1145/2185520.2185541}
\showDOI{\tempurl}


\bibitem[\protect\citeauthoryear{Si}{Si}{2015}]%
        {si2015tetgen}
\bibfield{author}{\bibinfo{person}{Hang Si}.} \bibinfo{year}{2015}\natexlab{}.
\newblock \showarticletitle{TetGen, a Delaunay-Based Quality Tetrahedral Mesh
  Generator}.
\newblock \bibinfo{journal}{\emph{ACM Trans. Math. Softw.}}
  \bibinfo{volume}{41}, \bibinfo{number}{2}, Article \bibinfo{articleno}{11}
  (\bibinfo{date}{Feb.} \bibinfo{year}{2015}), \bibinfo{numpages}{36}~pages.
\newblock
\showISSN{0098-3500}
\urldef\tempurl%
\url{https://doi.org/10.1145/2629697}
\showDOI{\tempurl}


\bibitem[\protect\citeauthoryear{Singh and Fiume}{Singh and Fiume}{1998}]%
        {wires}
\bibfield{author}{\bibinfo{person}{Karan Singh} {and} \bibinfo{person}{Eugene
  Fiume}.} \bibinfo{year}{1998}\natexlab{}.
\newblock \showarticletitle{Wires: {A} Geometric Deformation Technique}. In
  \bibinfo{booktitle}{\emph{Proceedings of the 25th Annual Conference on
  Computer Graphics and Interactive Techniques, {SIGGRAPH} 1998, Orlando, FL,
  USA, July 19-24, 1998}}, \bibfield{editor}{\bibinfo{person}{Steve
  Cunningham}, \bibinfo{person}{Walt Bransford}, {and}
  \bibinfo{person}{Michael~F. Cohen}} (Eds.). \bibinfo{publisher}{{ACM}},
  \bibinfo{address}{New York, NY, USA}, \bibinfo{pages}{405--414}.
\newblock
\urldef\tempurl%
\url{https://doi.org/10.1145/280814.280946}
\showDOI{\tempurl}


\bibitem[\protect\citeauthoryear{Sorkine and Cohen-Or}{Sorkine and
  Cohen-Or}{2004}]%
        {sorkine2004LSmeshes}
\bibfield{author}{\bibinfo{person}{Olga Sorkine} {and} \bibinfo{person}{Daniel
  Cohen-Or}.} \bibinfo{year}{2004}\natexlab{}.
\newblock \showarticletitle{Least-squares Meshes}. In
  \bibinfo{booktitle}{\emph{Proceedings of Shape Modeling International}}
  (Genova, Italy). \bibinfo{publisher}{IEEE Computer Society Press},
  \bibinfo{address}{Piscataway, NJ, USA}, \bibinfo{pages}{191--199}.
\newblock


\bibitem[\protect\citeauthoryear{Stam}{Stam}{2003}]%
        {stam2003flows}
\bibfield{author}{\bibinfo{person}{Jos Stam}.} \bibinfo{year}{2003}\natexlab{}.
\newblock \showarticletitle{Flows on Surfaces of Arbitrary Topology}. In
  \bibinfo{booktitle}{\emph{ACM SIGGRAPH 2003 Papers}} (San Diego, California)
  \emph{(\bibinfo{series}{SIGGRAPH ’03})}. \bibinfo{publisher}{Association
  for Computing Machinery}, \bibinfo{address}{New York, NY, USA},
  \bibinfo{pages}{724–731}.
\newblock
\showISBNx{1581137095}
\urldef\tempurl%
\url{https://doi.org/10.1145/1201775.882338}
\showDOI{\tempurl}


\bibitem[\protect\citeauthoryear{Stanculescu, Chaine, Cani, and
  Singh}{Stanculescu et~al\mbox{.}}{2013}]%
        {neobarok}
\bibfield{author}{\bibinfo{person}{Lucian Stanculescu},
  \bibinfo{person}{Rapha{\"e}lle Chaine}, \bibinfo{person}{Marie-Paule Cani},
  {and} \bibinfo{person}{Karan Singh}.} \bibinfo{year}{2013}\natexlab{}.
\newblock \showarticletitle{Sculpting Multi-dimensional Nested Structures}.
\newblock \bibinfo{journal}{\emph{Comput. Graph.-UK}} \bibinfo{volume}{37},
  \bibinfo{number}{6} (\bibinfo{date}{Oct.} \bibinfo{year}{2013}),
  \bibinfo{pages}{753--763}.
\newblock
\newblock
\shownote{Special issue: Shape Modeling International (SMI) Conference 2013.}


\bibitem[\protect\citeauthoryear{Surazhsky, Surazhsky, Kirsanov, Gortler, and
  Hoppe}{Surazhsky et~al\mbox{.}}{2005}]%
        {surazhsky2005fast}
\bibfield{author}{\bibinfo{person}{Vitaly Surazhsky}, \bibinfo{person}{Tatiana
  Surazhsky}, \bibinfo{person}{Danil Kirsanov}, \bibinfo{person}{Steven~J.
  Gortler}, {and} \bibinfo{person}{Hugues Hoppe}.}
  \bibinfo{year}{2005}\natexlab{}.
\newblock \showarticletitle{Fast Exact and Approximate Geodesics on Meshes}.
\newblock \bibinfo{journal}{\emph{ACM Trans. Graph.}} \bibinfo{volume}{24},
  \bibinfo{number}{3} (\bibinfo{date}{July} \bibinfo{year}{2005}),
  \bibinfo{pages}{553–560}.
\newblock
\showISSN{0730-0301}
\urldef\tempurl%
\url{https://doi.org/10.1145/1073204.1073228}
\showDOI{\tempurl}


\bibitem[\protect\citeauthoryear{Takayama, Panozzo, Sorkine-Hornung, and
  Sorkine-Hornung}{Takayama et~al\mbox{.}}{2013}]%
        {takayama2013sketch}
\bibfield{author}{\bibinfo{person}{Kenshi Takayama}, \bibinfo{person}{Daniele
  Panozzo}, \bibinfo{person}{Alexander Sorkine-Hornung}, {and}
  \bibinfo{person}{Olga Sorkine-Hornung}.} \bibinfo{year}{2013}\natexlab{}.
\newblock \showarticletitle{Sketch-based Generation and Editing of Quad
  Meshes}.
\newblock \bibinfo{journal}{\emph{ACM Trans. Graph.}} \bibinfo{volume}{32},
  \bibinfo{number}{4}, Article \bibinfo{articleno}{97} (\bibinfo{date}{July}
  \bibinfo{year}{2013}), \bibinfo{numpages}{8}~pages.
\newblock
\showISSN{0730-0301}
\urldef\tempurl%
\url{https://doi.org/10.1145/2461912.2461955}
\showDOI{\tempurl}


\bibitem[\protect\citeauthoryear{Thiel, Singh, and Balakrishnan}{Thiel
  et~al\mbox{.}}{2011}]%
        {thiel2011elasticurves}
\bibfield{author}{\bibinfo{person}{Yannick Thiel}, \bibinfo{person}{Karan
  Singh}, {and} \bibinfo{person}{Ravin Balakrishnan}.}
  \bibinfo{year}{2011}\natexlab{}.
\newblock \showarticletitle{Elasticurves: Exploiting Stroke Dynamics and
  Inertia for the Real-Time Neatening of Sketched 2D Curves}. In
  \bibinfo{booktitle}{\emph{Proceedings of the 24th Annual ACM Symposium on
  User Interface Software and Technology}} (Santa Barbara, California, USA)
  \emph{(\bibinfo{series}{UIST ’11})}. \bibinfo{publisher}{Association for
  Computing Machinery}, \bibinfo{address}{New York, NY, USA},
  \bibinfo{pages}{383–392}.
\newblock
\showISBNx{9781450307161}
\urldef\tempurl%
\url{https://doi.org/10.1145/2047196.2047246}
\showDOI{\tempurl}


\bibitem[\protect\citeauthoryear{Tong, Alliez, Cohen-Steiner, and Desbrun}{Tong
  et~al\mbox{.}}{2006}]%
        {tong2006designing}
\bibfield{author}{\bibinfo{person}{Yiying Tong}, \bibinfo{person}{Pierre
  Alliez}, \bibinfo{person}{David Cohen-Steiner}, {and}
  \bibinfo{person}{Mathieu Desbrun}.} \bibinfo{year}{2006}\natexlab{}.
\newblock \showarticletitle{Designing Quadrangulations with Discrete Harmonic
  Forms}. In \bibinfo{booktitle}{\emph{Proceedings of the Fourth Eurographics
  Symposium on Geometry Processing}} (Cagliari, Sardinia, Italy)
  \emph{(\bibinfo{series}{SGP ’06})}. \bibinfo{publisher}{Eurographics
  Association}, \bibinfo{address}{Goslar, DEU}, \bibinfo{pages}{201–210}.
\newblock
\showISBNx{3905673363}


\bibitem[\protect\citeauthoryear{Turk}{Turk}{2001}]%
        {turk2001texture}
\bibfield{author}{\bibinfo{person}{Greg Turk}.}
  \bibinfo{year}{2001}\natexlab{}.
\newblock \showarticletitle{Texture Synthesis on Surfaces}. In
  \bibinfo{booktitle}{\emph{Proceedings of the 28th Annual Conference on
  Computer Graphics and Interactive Techniques}}
  \emph{(\bibinfo{series}{SIGGRAPH ’01})}. \bibinfo{publisher}{Association
  for Computing Machinery}, \bibinfo{address}{New York, NY, USA},
  \bibinfo{pages}{347–354}.
\newblock
\showISBNx{158113374X}
\urldef\tempurl%
\url{https://doi.org/10.1145/383259.383297}
\showDOI{\tempurl}


\bibitem[\protect\citeauthoryear{Turquin, Wither, Boissieux, Cani, and
  Hughes}{Turquin et~al\mbox{.}}{2007}]%
        {turquin2007sketch}
\bibfield{author}{\bibinfo{person}{Emmanuel Turquin}, \bibinfo{person}{Jamie
  Wither}, \bibinfo{person}{Laurence Boissieux}, \bibinfo{person}{Marie-Paule
  Cani}, {and} \bibinfo{person}{John~F. Hughes}.}
  \bibinfo{year}{2007}\natexlab{}.
\newblock \showarticletitle{A Sketch-Based Interface for Clothing Virtual
  Characters}.
\newblock \bibinfo{journal}{\emph{IEEE Comput. Graph. Appl.}}
  \bibinfo{volume}{27}, \bibinfo{number}{1} (\bibinfo{date}{Jan.}
  \bibinfo{year}{2007}), \bibinfo{pages}{72--81}.
\newblock
\showISSN{0272-1716}
\urldef\tempurl%
\url{https://doi.org/10.1109/MCG.2007.1}
\showDOI{\tempurl}


\bibitem[\protect\citeauthoryear{Wesche and Seidel}{Wesche and Seidel}{2001}]%
        {wesche2001freedrawer}
\bibfield{author}{\bibinfo{person}{Gerold Wesche} {and}
  \bibinfo{person}{Hans-Peter Seidel}.} \bibinfo{year}{2001}\natexlab{}.
\newblock \showarticletitle{FreeDrawer: A Free-form Sketching System on the
  Responsive Workbench}. In \bibinfo{booktitle}{\emph{Proceedings of the ACM
  symposium on Virtual reality software and technology}}.
  \bibinfo{publisher}{ACM}, \bibinfo{address}{New York, NY, USA},
  \bibinfo{pages}{167--174}.
\newblock


\bibitem[\protect\citeauthoryear{Wiese, Israel, Meyer, and Bongartz}{Wiese
  et~al\mbox{.}}{2010}]%
        {wiese2010investigating}
\bibfield{author}{\bibinfo{person}{Eva Wiese}, \bibinfo{person}{Johann~Habakuk
  Israel}, \bibinfo{person}{Achim Meyer}, {and} \bibinfo{person}{Sara
  Bongartz}.} \bibinfo{year}{2010}\natexlab{}.
\newblock \showarticletitle{Investigating the Learnability of Immersive
  Free-Hand Sketching}. In \bibinfo{booktitle}{\emph{Proceedings of the Seventh
  Sketch-Based Interfaces and Modeling Symposium}} (Annecy, France)
  \emph{(\bibinfo{series}{SBIM ’10})}. \bibinfo{publisher}{Eurographics
  Association}, \bibinfo{address}{Goslar, DEU}, \bibinfo{pages}{135–142}.
\newblock
\showISBNx{9783905674255}


\bibitem[\protect\citeauthoryear{Xing, Nagano, Chen, Xu, Wei, Zhao, Lu, Kim,
  and Li}{Xing et~al\mbox{.}}{2019}]%
        {xing2019hairbrush}
\bibfield{author}{\bibinfo{person}{Jun Xing}, \bibinfo{person}{Koki Nagano},
  \bibinfo{person}{Weikai Chen}, \bibinfo{person}{Haotian Xu},
  \bibinfo{person}{Li-yi Wei}, \bibinfo{person}{Yajie Zhao},
  \bibinfo{person}{Jingwan Lu}, \bibinfo{person}{Byungmoon Kim}, {and}
  \bibinfo{person}{Hao Li}.} \bibinfo{year}{2019}\natexlab{}.
\newblock \showarticletitle{HairBrush for Immersive Data-Driven Hair Modeling}.
  In \bibinfo{booktitle}{\emph{Proceedings of the 32nd Annual ACM Symposium on
  User Interface Software and Technology}} (New Orleans, LA, USA)
  \emph{(\bibinfo{series}{UIST ’19})}. \bibinfo{publisher}{Association for
  Computing Machinery}, \bibinfo{address}{New York, NY, USA},
  \bibinfo{pages}{263–279}.
\newblock
\showISBNx{9781450368162}
\urldef\tempurl%
\url{https://doi.org/10.1145/3332165.3347876}
\showDOI{\tempurl}


\bibitem[\protect\citeauthoryear{Xu, Chang, Sheffer, Bousseau, McCrae, and
  Singh}{Xu et~al\mbox{.}}{2014}]%
        {xu2014true2form}
\bibfield{author}{\bibinfo{person}{Baoxuan Xu}, \bibinfo{person}{William
  Chang}, \bibinfo{person}{Alla Sheffer}, \bibinfo{person}{Adrien Bousseau},
  \bibinfo{person}{James McCrae}, {and} \bibinfo{person}{Karan Singh}.}
  \bibinfo{year}{2014}\natexlab{}.
\newblock \showarticletitle{True2Form: 3D Curve Networks from 2D Sketches via
  Selective Regularization}.
\newblock \bibinfo{journal}{\emph{ACM Trans. Graph.}} \bibinfo{volume}{33},
  \bibinfo{number}{4}, Article \bibinfo{articleno}{131} (\bibinfo{date}{July}
  \bibinfo{year}{2014}), \bibinfo{numpages}{13}~pages.
\newblock
\showISSN{0730-0301}
\urldef\tempurl%
\url{https://doi.org/10.1145/2601097.2601128}
\showDOI{\tempurl}


\end{thebibliography}
